# JEDI: These aren't the JSON documents you're looking for... (Extended Version*)

Thomas Hütter
Nikolaus Augsten
thomas.huetter@plus.ac.at
nikolaus.augsten@plus.ac.at
University of Salzburg
Austria

Christoph M. Kirsch
ck@cs.uni-salzburg.at
University of Salzburg
Austria
Czech Technical University
Czech Republic

Michael J. Carey
Chen Li
mjcarey@ics.uci.edu
chenli@ics.uci.edu
University of California, Irvine
USA

## ABSTRACT

The JavaScript Object Notation (JSON) is a popular data format used in document stores to natively support semi-structured data. In this paper, we address the problem of JSON similarity lookup queries: given a query document and a distance threshold $\tau$, retrieve all JSON documents that are within $\tau$ from the query document. Due to its recursive definition, JSON data are naturally represented as trees. Different from other hierarchical formats such as XML, JSON supports both ordered and unordered sibling collections within a single document. This feature poses a new challenge to the tree model and distance computation. We propose JSON tree, a lossless tree representation of JSON documents, and define the JSON Edit Distance (JEDI), the first edit-based distance measure for JSON documents. We develop an algorithm, called QuickJEDI, for computing JEDI by leveraging a new technique to prune expensive sibling matchings. It outperforms a baseline algorithm by an order of magnitude in runtime. To boost the performance of JSON similarity queries, we introduce an index called JSIM and a highly effective upper bound based on tree sorting. Our algorithm for the upper bound runs in $O(n\tau)$ time and $O(n + \tau \log n)$ space, which substantially improves the previous best bound of $O(n^2)$ time and $O(n \log n)$ space (where $n$ is the tree size). Our experimental evaluation shows that our solution scales to databases with millions of documents and JSON trees with tens of thousands of nodes.

## 1 INTRODUCTION

The JavaScript Object Notation (JSON) has evolved into one of the most prominent data formats. It is used in a large variety of scenarios, e.g., to publish datasets [21, 54] or as an open-standard data interchange format in mobile and web applications [20]. JSON-like formats are also used in document stores to natively support semi-structured data [1, 6, 39].

As JSON does not enforce a schema, it increases the variety in which a given piece of information can be represented. Consider the following scenario: a web crawler collects movies in JSON format from multiple sources and saves them in a document store. In order to avoid duplicate entries, the crawler queries the database for the existence of the movie to be inserted. However, a query for exact duplicates is not effective since key names or the structure will typically vary. Consider the document in Figure 1a, which was discovered by the crawler. The document in Figure 1b already resides in the database. Both documents represent the same movie, but due to their different representations a query for exact duplicates will fail. Detecting near-duplicate entries remains a challenge.

```
{
    "title" : "Star Wars -
        A New Hope",
    "running time" : 125,
    "cast" : {
        "Han" : "Ford",
        "Leia" : "Fisher"
    }
}
        (a)
```

```
{
    "cast" : [
        "Ford",
        "Fisher"
    ],
    "running time" : 125,
    "name" : "Star Wars -
        A New Hope",
}
        (b)
```

Figure 1: Two JSON representations of the same movie.

In this paper, we study the following problem of similarity queries for JSON: given a query document $T_q$, retrieve all documents $T_i$ from a database of JSON documents that are within a given distance threshold $\tau$ from $T_q$. To answer such queries, a distance function that assesses the similarity of two JSON documents is needed.

JSON similarity queries are poorly supported in many existing systems. Stand-alone tools for JSON differences are either line-based hence ignore the hierarchical information [16, 26], or do not provide any guarantees on the quality of the result [12]. Similarity functions related to JSON in database systems are limited to basic values, e.g., strings and sets, and cannot be used to compute the distance between JSON documents [18, 33, 35, 39, 42]. For other data formats, similarity queries and the related distance measures are well studied, e.g., for XML [17, 22, 32, 37] data, which – like JSON – is a hierarchical data format. Common approaches for XML are based on the well-known tree edit distance [41], which is the minimal difference between two documents respecting both their hierarchical structure and data values. For JSON, assessing a minimal, edit-based difference is still an unsolved problem.

Assessing the edit-based difference between JSON documents is challenging. JSON differs from other hierarchical data formats, such as XML, in that it supports both ordered and unordered sibling nodes within a single document. This uniqueness calls (1) for a tree representation that models both types of siblings, and (2) for a distance function that assesses the similarity of the resulting trees. In particular, the support for unordered sibling collections poses a computational challenge: We show that computing the minimal difference between JSON documents is NP-hard when no

---

restrictions are imposed on the standard set of node edit operations, i.e., insertion, deletion, and renaming.

We solve the problem of computing a minimal, edit-based difference between JSON documents. (1) We develop a lossless tree representation of JSON that models both ordered and unordered siblings. (2) We show that the edit distance in its general formulation leads to non-intuitive results. We therefore restrict the edit operations to respect the nested document structure of JSON and propose the first edit-based distance measure for JSON documents, called *JSON Edit Distance (JEDI)*. The function guarantees that the difference is minimal and the document nesting is respected.

We present an algorithm for JEDI, called *QuickJEDI*, which is based on a recursive solution. Compared to previous edit distance algorithms for related problems [60, 61], this algorithm leverages a novel technique, the aggregate size bound, to prune the expensive min-cost matching between sibling sets in each recursive step. This optimization leads to runtime improvements of up to an order of magnitude. We further propose the *JSIM* index that only searches the $\tau$-range around the query document instead of scanning all documents. JSIM is a 4-level tree and each level routes the search into one or more branches. A new technique allows us to reduce the $\tau$-range at each level, thus reducing the total number of explored branches. The documents returned by the index are filtered with a highly effective upper bound based on tree sorting, for which we improve the computational complexity from quadratic to linear.

The main contributions of this paper are:

- We show that the existing formulation of the tree edit distance can lead to non-intuitive results and is NP-hard for JSON. To solve the problem, we introduce JSON trees, a lossless tree representation of JSON documents, and JEDI, the first edit-based distance measure for JSON.
- We develop a new algorithm, QuickJEDI, for computing JEDI in $O(n^2 d \log d)$ time and $O(n^2)$ space for JSON trees of size $n$ and maximum degree $d$. The algorithm leverages the new aggregate size bound to prune expensive sibling matchings.
- To improve the performance of JSON similarity queries, we introduce (1) a novel index called JSIM; (2) an effective upper bound based on tree sorting, and an algorithm for computing the bound in $O(n\tau)$ time and $O(n + \tau \log n)$ space, which substantially improves the previous best bound of $O(n^2)$ time and $O(n \log n)$ space.
- Our empirical study on 22 JSON datasets suggests that our solution scales to databases with millions of documents and can handle large JSON trees with tens of thousands of nodes.

## 2 EDIT-BASED DISTANCE FOR JSON TREES

We now introduce a new distance measure that assesses the minimal difference of two JSON documents by a given set of allowable edit operations and a novel tree representation of JSON data. To our best knowledge, this measure is the first distance for JSON that respects its nested structure and provides quality guarantees, i.e., the difference defining the distance is guaranteed to be minimal.

*The JSON Data Format.* We recap the definition of the JSON data format (cf. RFC8259 [8]). A JSON document is recursively composed of values, arrays, and objects: (1) A *value* is either a literal (string, number, boolean, or null), an object, or an array. (2) An *array* is an ordered, possibly empty list of *values* enclosed by brackets. (3) An *object* is an unordered, possibly empty collection of key-value pairs enclosed by curly braces. The *keys* (called "names" in [8]) are string literals that are unique within an object.

EXAMPLE 1. *The JSON document in Figure 1a is an object of three key-value pairs. The keys are* `"title"`, `"running time"`, *and* `"cast"`. *The value of* `"cast"` *is an object, and the other values are string and number literals.*

### 2.1 JSON Tree Representation

Due to its recursive definition, JSON is hierarchically structured and naturally represented as a tree. The specifics of transforming a JSON document into a tree, however, are not obvious. Previous attempts to model JSON as trees are unsuitable for distances based on a minimal number of node edit operations because either (1) the object and array information is not modelled [34, 46], e.g., `[['A']]` and `'A'` are transformed to identical trees such that the structural information is lost; or (2) arrays are modeled as objects with the array index as a key [7, 47], which generates an error of $O(n)$ when a single element in an array of size $n$ is missing. Consider two arrays `['A', 'B', 'C', 'D']` and `['B', 'C', 'D']`: the array index keys of all identical elements differ due to element `'A'` that is not present in the second array. Tree models for XML documents are not suitable since XML is ordered by definition; although XML has been modeled as unordered trees to capture the semantics of data-centric XML [2], these models do not support a mix of ordered and unordered siblings.

*JSON Tree.* We introduce the new concept of a *JSON tree*. The constraints that we impose on JSON trees model all aspects of JSON data and allow for a lossless transformation between JSON documents and JSON trees.

A JSON tree $T = (N, E, \Lambda, \Psi, <_S)$ is a tree with nodes $N$ and edges $E \subseteq N \times N$. The label of node $v$, $\Lambda(v)$, is a literal value; the labels of array and object nodes are *null*. Function $\Psi$ assigns a type to each node $v \in N$, $\Psi(v) \in \{\text{object}, \text{array}, \text{key}, \text{literal}\}$. The *sibling order*, $<_S$, defines a strict, partial order on the nodes of a tree. Two nodes $x, y \in N(T)$ of a JSON tree are *comparable*, i.e., $x <_S y$ or $y <_S x$, iff one of the following holds:

(1) $x$ and $y$ are children of the same array node; or
(2) there is an ancestor $x'$ of $x$ (including $x$) and an ancestor $y'$ of $y$ (including $y$) such that $x'$ and $y'$ are comparable.

In the second condition, $x <_S y$ iff $x' <_S y'$. Intuitively, the order among the children of an array node imposes an order on the subtrees rooted in these children; all other nodes are incomparable. The children of an object node (i.e., key nodes) must have unique labels among their siblings.

*Transformation.* A JSON document is transformed into a JSON tree by recursively unnesting the document. Objects become nodes of type object (label *null*) with key node children; a key node (labeled with its name) has a single child subtree that represents its value; an array (label *null*) becomes an array node with the $i$-th value in its list becoming the $i$-th child subtree defining the sibling order $<_S$. Literals are leaf nodes of type literal (labeled with the respective literal value).

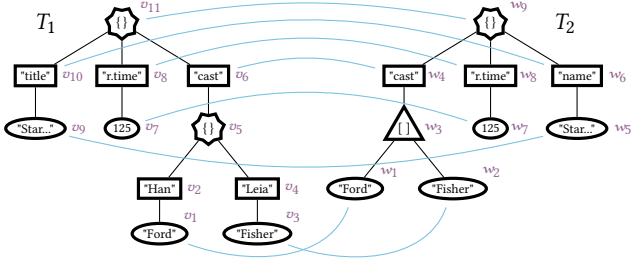

**Figure 2: JSON trees of the documents in Figure 1. Object nodes are visualized as stars with symbol { }, array nodes as triangles with symbol [ ], keys as rectangles, and literals as ellipses with their original labels, respectively. Blue lines depict the JSON edit mapping, and $v_i$, $w_j$ are node identifiers.**

EXAMPLE 2. *Figure 2 shows the JSON tree representation of the two documents in Figure 1.*

*Notation.* With $N(T)$ resp. $E(T)$, we denote the nodes resp. edges of a JSON tree $T$. $|T| = |N(T)|$ is the size of $T$, $v \in T$ is shorthand for $v \in N(T)$. The parent of a node $v \in T$ is $p(v)$, the set of its children is $chd(v)$, the degree is $deg(v) = |chd(v)|$; the *degree* of tree $T$, $deg(T)$, is the largest degree of a node in $T$; $anc(v)$ and $desc(v)$ denote the set of ancestors resp. descendants of $v$ (excluding $v$). The lowest common ancestor of two nodes $v, w$ is $lca(v, w)$.

$T[v]$ denotes the subtree rooted in node $v$. The subforest of node $v$, denoted $F[v]$, is the set of subtrees of its children, $T[v_i], v_i \in chd(v)$. If an order is defined on $chd(v)$, then the subforest $F[v]$ is ordered by the root nodes of its subtrees. We use $\epsilon$ to denote the *empty node*, which is not part of any tree. We define $T[\epsilon]$ to be the *empty subtree* with $N(T[\epsilon]) = \emptyset$ and $E(T[\epsilon]) = \emptyset$, and $F[\epsilon] = \emptyset$ to be the *empty subforest*. The *postorder* traversal recursively visits all children of a node $v$ before visiting $v$ (ordered children in ascending order and unordered children in arbitrary order; $post(v)$ is the position of node $v$ in a given postorder traversal.

## 2.2 JSON Edit Distance (JEDI)

Given two JSON trees, our goal is to assess their similarity. We aim for a similarity measure that captures fine-grained differences, allows an intuitive interpretation of the similarity value, and guarantees the minimality of the similarity value. A well-known approach that satisfies these requirements is the edit distance, which has been applied to strings [58], trees [41], and graphs [24].

The edit distances for general, rooted, labeled trees [48, 62], however, are not applicable to JSON trees since they can only deal with *either* ordered *or* unordered trees, but not with a mix of the two. In JSON trees, the order of array children must be respected, whereas the order of object children must be ignored. Note that all nodes in subtree $T[c_i]$ appear before the nodes in $T[c_j]$ if $c_i <_S c_j$, i.e., the order imposed by an array is propagated to the subtrees rooted in the children. We are the first to define an edit distance that can deal with both ordered and unordered siblings in a single tree. Further, we respect the node types of JSON, e.g., a literal value should not be aligned to a key node.

Similar to the edit distance for other data types, we define the JSON edit distance (JEDI) as the minimum number of edit operations required to transform one tree to the other. Allowable operations include: *delete* node $v$ and connect its children to the parent of $v$; *insert* a new node $w$ between an existing node $v$ and a possibly empty subset of $v$'s children; and *rename* the label of node $v$.

*JSON Edit Mapping.* Following previous works, we formally define the JSON edit distance using the concept of an edit mapping. The edit mapping aligns the nodes of the input trees, $T_1$ and $T_2$, and must respect some constraints to be valid. The interpretation is as follows: nodes in $T_1$ that are not mapped are deleted, nodes in $T_2$ that are not mapped are inserted, and nodes that are mapped are renamed. The constraints imposed on the mapping control which edit operations are allowable depending on the tree context; they are discussed in detail below.

DEFINITION 1 (JSON EDIT MAPPING). *A mapping $M \subseteq N(T_1) \times N(T_2)$ is a JSON edit mapping from $T_1$ to $T_2$ iff the following constraints hold for any node pairs $(v, w), (v', w'), (v'', w'') \in M$:*
  (1) $v = v'$ iff $w = w'$ [one-to-one],
  (2) $v$ is an ancestor of $v'$ iff $w$ is an ancestor of $w'$ [ancestor],
  (3) $type(v) = type(w)$ [type],
  (4) *if* $v <_S v'$ *and* $w$ *is comparable to* $w'$ *in* $<_S$, *then* $w <_S w'$ [array-order],
  (5) $lca(v, v')$ *is a proper ancestor of* $v''$ *iff* $lca(w, w')$ *is a proper ancestor of* $w''$ [document-preserving].

*A mapping $M' \subseteq M$ between two subforest $F_1[v]$ and $F_2[w]$ is an edit mapping iff $M'$ is an edit mapping from $T_1[v]$ to $T_2[w]$.*

The cost of all edit operations is one except for rename: if the labels of the mapped nodes are identical, then the cost is zero. The cost of an edit mapping, $\gamma(M)$, is the total cost of all edit operations. The edit distance is defined as the cost of the edit mapping with the lowest cost.

EXAMPLE 3. *Figure 2 shows a JSON edit mapping between two JSON trees. The cost of the mapping is 5: delete nodes* "Han", "Leia", *and* {} *from the left tree; insert* [] *into the right tree; rename* "title" *to* "name". *There is no mapping with a lower cost, thus JEDI is 5.*

Constraints (1) and (2) of the edit mapping ensure that the node mapping can be interpreted as a set of edit operations. Constraint (3) ensures that labels can only be renamed between nodes of the same type. This prevents that nodes with identical labels but different types (e.g., a key and a literal value may have identical labels) are mapped at zero cost, thus ignoring their difference. Note that it is still possible to substitute (delete and insert) a node of one type by a node of another type, but the cost is higher than for rename. The array-order constraint (4) uses the partial order $<_S$ defined on JSON trees to enforce the order imposed by array nodes; children of object nodes are not restricted and can be arbitrarily mapped.

*Document-Preserving Constraint.* The recursive definition of JSON gives rise to its nested document structure. A nested document (e.g., representing the cast of a movie) often is meaningful only in the context of the enclosing document (in the example, the movie the cast belongs to). Constraint (5), the *document-preserving* constraint, forces the edit mappings to respect the nested document structure of JSON and leads to more intuitive mappings. In particular, shortcuts that delete the root nodes of subtrees (thus disassembling the documents they root), rearrange their nested subtrees, and recompose the nested subtrees into new documents by inserting new subtree roots are prevented. We illustrate the effect of this constraint in Example 4.

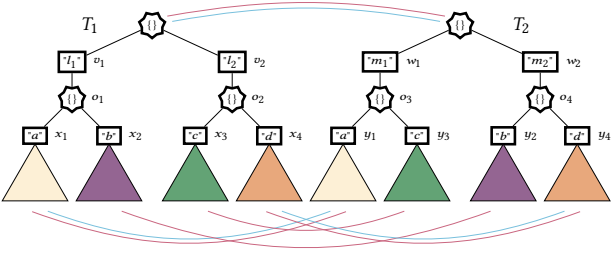

Figure 3: Edit mappings with and without the document-preserving constraint.

EXAMPLE 4. *Consider the schematic illustration of the two JSON trees $T_1$ and $T_2$ in Figure 3. Each tree consists of two subtrees $T_1[v_1]$, $T_1[v_2]$ resp. $T_2[w_1]$, $T_2[w_2]$, which in turn are composed of an object node and two smaller subtrees $T_1[x_i]$ resp. $T_2[y_j]$ each, $i, j \leq 4$. The subtree pairs $T_1[x_i]$ and $T_2[y_j]$, $i = j$, are identical (same color in the figure) and are all of size n. Any two subtrees $T_1[v_i]$ and $T_2[w_j]$, $i \neq j$, are different with an edit mapping of cost $O(n)$.*

*The minimum-cost edit mapping (with document-preserving constraint) will delete $T_1[x_2]$ and $T_1[x_3]$, and insert their identical counterparts $T_2[y_2]$ and $T_2[y_3]$ since they belong to different documents in $T_2$. An edit mapping that does not respect the document-preserving constraint, however, has only cost 8: delete nodes $v_1, v_2, o_1, o_2$, insert $o_3$ as parent of $y_1, y_3$; $o_4$ as parent of $y_2, y_4$; $w_1$ as parent of $o_3$; and $w_2$ as parent of $o_4$. Without the document-preserving constraint, rearranged subtrees form new documents, which is not desired for JSON trees.*

As a pleasant side effect, the document-preserving constraint substantially reduces the search space for the minimal cost mapping and allows for faster algorithms. In fact, we show that finding a minimum JSON edit mapping that ignores the document-preserving constraint is an NP-hard problem. The proof is by reducing the problem of exact cover by 3-sets (*X3C*).

THEOREM 1. *Without the document-preserving constraint, the problem of computing the JSON edit distance between two JSON trees is NP-hard.*

PROOF. Cf. Section A.1. □

## 3 AN EFFICIENT ALGORITHM FOR JEDI

Next, we introduce QuickJEDI, an efficient algorithm for computing the JSON edit distance. We first discuss a baseline solution, analyze its performance bottlenecks, and finally propose effective techniques to address these bottlenecks.

### 3.1 A Baseline Algorithm

None of the previous algorithms that computes the minimum edit distances between trees is applicable in our scenario due to the type and the array-order constraints in the JSON edit mapping (cf. Definition 1). Our baseline extends two algorithms for the so-called *constrained tree edit distance*. These algorithms compute minimal edits under the document-preserving constraint (constraint (5) in the JSON edit mapping) for ordered [60] resp. unordered trees [61]. Since a single JSON tree may include both ordered and unordered siblings, neither of the two algorithms is applicable; also, these algorithms deal with generic trees and do not consider node types.

We recap the solutions for the constrained tree edit distance and show how they can be extended to compute the JSON edit distance. Both algorithms are based on a recursive solution that is implemented using dynamic programming.

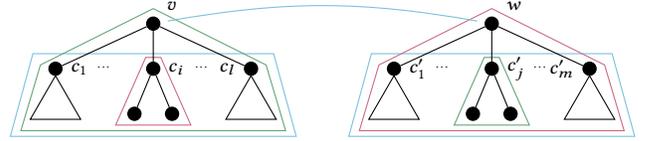

Figure 4: Recursive decomposition of two trees; pairs of subtrees resp. subforests of the same color form the subproblems required to compute the distance btw. *T[v]* and *T[w]*.

*Recursive Solution.* The recursive solution decomposes two trees $T_1$ and $T_2$ with root nodes $v \in T_1$ and $w \in T_2$ into subtrees and subforests as illustrated in Figure 4. The distance between $T_1$ and $T_2$ is computed from the distances between the subproblems resulting from their decomposition. With dt($v, w$) we denote the *tree distance* between subtrees $T_1[v]$ and $T_2[w]$, and df($v, w$) denotes the *forest distance* between subforests $F_1[v]$ and $F_2[w]$. Then, the recursive solution is defined as follows:

df($\epsilon, \epsilon$) = 0; dt($\epsilon, \epsilon$) = 0

$$\text{df}(v, \epsilon) = \sum_{c \in \text{chd}(v)} \text{dt}(c, \epsilon); \quad \text{dt}(v, \epsilon) = \text{df}(v, \epsilon) + \gamma(v, \epsilon) \quad (1)$$

$$\text{df}(\epsilon, w) = \sum_{c' \in \text{chd}(w)} \text{dt}(\epsilon, c'); \quad \text{dt}(\epsilon, w) = \text{df}(\epsilon, w) + \gamma(\epsilon, w)$$

$$\text{df}(v, w) = \min \begin{cases} \text{df}(\epsilon, w) + \min_{c' \in \text{chd}(w)}\{\text{df}(v, c') - \text{df}(\epsilon, c')\} & (2a) \\ \text{df}(v, \epsilon) + \min_{c \in \text{chd}(v)}\{\text{df}(c, w) - \text{df}(c, \epsilon)\} & (2b) \\ \text{Min-cost-matching (chd}(v), \text{chd}(w)) & (2c) \end{cases}$$

$$\text{dt}(v, w) = \min \begin{cases} \text{dt}(\epsilon, w) + \min_{c' \in \text{chd}(w)}\{\text{dt}(v, c') - \text{dt}(\epsilon, c')\} & (3a) \\ \text{dt}(v, \epsilon) + \min_{c \in \text{chd}(v)}\{\text{dt}(c, w) - \text{dt}(c, \epsilon)\} & (3b) \\ \text{df}(v, w) + \gamma(v, w) & (3c) \end{cases}$$

The tree distance, dt($v, w$), is the minimum cost of three scenarios (cf. Figure 4 and Eq. 3), each of which represents an edit operation: (3a) $w$ is *inserted*, hence the nodes in subtree $T_1[v]$ are mapped to the nodes of one of $w$'s children $T_2[c'_j]$ (green), (3b) $v$ is *deleted*, hence the nodes of subtree $T_2[w]$ are mapped to the nodes of one of $v$'s children $T_1[c_i]$ (red), and (3c) $v$ is mapped to $w$ with *rename* cost $\gamma(v, w)$, hence also the subtrees of their children are mapped (blue); $\gamma(v, \epsilon)$ and $\gamma(\epsilon, w)$ denote the cost of deleting resp. inserting a node. The cost of matching the children of nodes $v$ and $w$ in scenario (3c) is equivalent to their forest distance $df(v, w)$. The *base cases* of the recursion are shown in Eq. 1. The forest distance, $df(v, w)$, (cf. Eq. 2) is computed analogously for insertion (2a) and deletion (2b). In the third scenario (2c), a minimum-cost matching between the subtrees rooted in $chd(v)$ and $chd(w)$ is established.

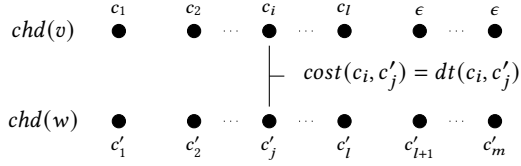

Figure 5: Bipartite graph for the nodes $chd(v)$ and $chd(w)$.

The *minimum-cost matching* $\mathcal{M}$ is one-to-one and models the subtrees rooted in $chd(v)$ and $chd(w)$ as nodes of a bipartite graph (cf. Figure 5); the cost of an edge between two subtrees rooted in $c_i \in chd(v)$ and $c'_j \in chd(w)$ is their tree distance, $dt(c_i, c'_j)$. In the unordered case [61], the minimum-cost bipartite graph matching, $M_{BPM(v,w)}$, with cost $\gamma(M_{BPM(v,w)}) = BPM(v, w)$ must be computed (e.g., using a min-cost max-flow algorithm [50]). In the ordered case [60], the subtree sequence edit distance matching, $M_{SED(v,w)}$, with cost $\gamma(M_{SED(v,w)}) = SED(v, w)$ must be computed.

*Adaption to JSON.* In Lemma 1, we show how previous solutions can be extended to compute JEDI between two JSON trees.

LEMMA 1. *Given two JSON trees $T_1$ and $T_2$, the recursive formulas (1), (2), and (3), compute the JSON edit distance between $T_1$ and $T_2$, $JEDI(T_1, T_2) = dt(root(T_1), root(T_2))$ with the following extensions:*

(1) *The minimum-cost matching $\mathcal{M} \subseteq chd(v) \times chd(w)$ observes the node type:*

$$\mathcal{M} = \begin{cases} M_{SED(v,w)} & \text{if } type(v) = type(w) = array \\ M_{BPM(v,w)} & \text{otherwise} \end{cases} \quad (4)$$

(2) *The rename cost must be redefined as follows:*

$$\gamma'(v, w) = \begin{cases} \gamma(v, w) & \text{if } type(v) = type(w) \\ \gamma(v, \epsilon) + \gamma(\epsilon, w) & \text{otherwise} \end{cases} \quad (5)$$

PROOF. Given a node pair $(v, w)$ with $v \in T_1$ and $w \in T_2$. We distinguish three cases: (1) *Both $v$ and $w$ are array nodes:* Proof as for the ordered, constrained edit distance [60]. (2) *Both $v$ and $w$ are non-array nodes:* Proof as for the unordered, constrained edit distance [61]. (3) *Otherwise:* W.l.o.g., we assume that $v$ is an array node (siblings ordered) and $w$ is a non-array node (siblings not ordered). The min-cost matching $\mathcal{M}$ between $chd(v)$ and $chd(w)$ must be computed using bipartite graph matching since the children of node $w$ can be arbitrarily rearranged to allow any matching without violating the order on $chd(v)$. — The rename cost for non-matching node types is the cost of deleting $v$ and inserting $w$ instead. □

*Dynamic Programming Implementation.* Algorithm 1 implements the recursive solution of Lemma 1. The results for subproblems are stored in two matrices, dt and df, each of size $(|T_1| + 1) \times (|T_2| + 1)$. The distance between subtrees $T_1[v]$ and $T_2[w]$ is stored in row $v$ and column $w$, and we refer to the value as $dt(v, w)$; similarly,

$df(v, w)$ stores the distance between subforests $F_1[v]$ and $F_2[w]$. Table 1 shows examples of a forest and a tree distance matrix.

*Initialization:* The first row and column of each matrix are initialized in lines 1-8. Mapping two empty trees has cost 0; for all other nodes, the cost results from summing up the deletion resp. insertion costs of their child subtrees, e.g., $dt(v_6, \epsilon) = 6$ (cf. Table 1) is the cost of deleting the subtree of node "cast" in Figure 2.

*Distance Computation:* The algorithm processes the tree nodes bottom-up in postorder and the distance matrices are filled row by row. We label the three cases in Eq. 2 (forest distance) with *insF* (2a), *delF* (2b), and *renF* (2c); the cases in Eq. 3 (tree distance) are labeled *insT* (3a), *delT* (3b), *renT* (3c). Due to the postorder traversal, all values required to compute $dt(v, w)$ and $df(v, w)$ are available in the distance matrices. To compute *renF*, a min-cost matching $\mathcal{M}$ among the children must be established. If both $v$ and $w$ are array nodes (ordered case), the edit distance between ordered sequences of siblings establishes the min-cost matching (line 16), in all other cases a bipartite graph matching must be computed (line 18). The distance between $T_1$ and $T_2$ results in the lower right corner of the tree distance matrix, e.g., $dt(root(T_1), root(T_2)) = 5$ in Table 1.

*Complexity:* The *space complexity* is dominated by the distance matrices of size $O(|T_1||T_2|)$. The runtime is dominated by the bipartite graph matching, which for a node pair $v, w$ with degrees $d_v = deg(v)$ and $d_w = deg(w)$ is computed in time $O(d_v \times d_w \times (d_v + d_w) \times log(d_v + d_w))$ using a min-cost max-flow algorithm [50]. For the overall algorithm (cf. Algorithm 1), the *runtime complexity* is $O(|T_1| \times |T_2| \times (deg(T_1) + deg(T_2)) \times log(deg(T_1) + deg(T_2)))$ [61].

| df | $\epsilon$ | $w_1$ | $w_2$ | ... | $w_6$ | $w_9$ | | dt | $\epsilon$ | $w_1$ | $w_2$ | ... | $w_6$ | $w_9$ |
|---|---|---|---|---|---|---|---|---|---|---|---|---|---|---|
| $\epsilon$ | 0 | 0 | 0 | ... | 1 | 8 | | $\epsilon$ | 0 | 1 | 1 | ... | 2 | 9 |
| $v_9$ | 0 | 0 | 0 | ... | 1 | 8 | | $v_9$ | 1 | 1 | 1 | ... | 1 | 8 |
| $v_{10}$ | 1 | 1 | 1 | ... | 0 | 7 | | $v_{10}$ | 2 | 2 | 2 | ... | 1 | 8 |
| ⋮ | ⋮ | ⋮ | ⋮ | ⋱ | ⋮ | ⋮ | | ⋮ | ⋮ | ⋮ | ⋮ | ⋱ | ⋮ | ⋮ |
| $v_6$ | 5 | 5 | 5 | ... | 5 | 8 | | $v_6$ | 6 | 5 | 5 | ... | 6 | 8 |
| $v_{11}$ | 10 | 10 | 10 | ... | 9 | 5 | | $v_{11}$ | 11 | 10 | 10 | ... | 10 | 5 |

Table 1: Forest and tree distance matrices df and dt for the JSON trees in Figure 2.

## 3.2 Avoiding the Expensive Min-Cost Matching

JEDI must compute the min-cost matching between the child subtrees of each node pair of the input trees. This step is expensive and dominates the overall runtime. In this section, we show that the expensive min-cost computation can be avoided in many cases, thus substantially improving the runtime of the distance computation.

The key idea is that the min-cost matching in Eq. (2) is the minimum of three values. Two of them are efficient to compute, one is the expensive matching. If we can show that the cost of the matching is higher than one of the other two values, the exact cost of the matching is irrelevant and the computation can be skipped.

We are the first to follow this approach. The challenge is to identify a lower bound on the min-cost matching that is both effective and can be computed efficiently. Efficiency is crucial since the lower bound filter will be evaluated *in addition* to the min-cost matching whenever the filter cannot avoid the matching computation. The min-cost matching is a bipartite graph matching in the unordered

---

**Algorithm 1:** JEDI-baseline($T_1, T_2$)

   **Input:** JSON trees $T_1$ and $T_2$.
   **Result:** JSON Edit Distance: JEDI($T_1, T_2$).
   /* Initialization. */
1  dt(0, 0) = 0   /* Tree distance matrix of size $T_1 + 1 \times T_2 + 1$. */
2  df(0, 0) = 0 /* Forest distance matrix of size $T_1 + 1 \times T_2 + 1$. */
3  **for** $v$ **in** $N(T_1)$ **do**
4     df($v$, 0) = $\sum_{c \in chd(v)}$ dt($c$, 0)
5     dt($v$, 0) = df($v$, 0) + $\gamma(v, \lambda)$
6  **for** $w$ **in** $N(T_2)$ **do**
7     df(0, $w$) = $\sum_{c' \in chd(w)}$ dt(0, $c'$)
8     dt(0, $w$) = df(0, $w$) + $\gamma(\lambda, w)$
   /* Distance computation. */
9  **for** $v$ **in** $N(T_1)$ **do**                                /* In postorder. */
10    **for** $w$ **in** $N(T_2)$ **do**                            /* In postorder. */
        /* Cost for inserting node $w$. */
11      insF = df(0, $w$) + $\min_{c' \in chd(w)}$ {df($v,c'$) - df(0,$c'$)}
12      insT = dt(0, $w$) + $\min_{c' \in chd(w)}$ {dt($v,c'$) - dt(0,$c'$)}
        /* Cost for deleting node $v$. */
13      delF = df($v$, 0) + $\min_{c \in chd(v)}$ {df($c,w$) - df($c$,0)}
14      delT = dt($v$, 0) + $\min_{c \in chd(v)}$ {dt($c,w$) - dt($c$,0)}
        /* Cost for renaming node $v$ to node $w$. */
15      **if** $type(v) == type(w) == array$ **then**
16         renF = SED($v$, $w$)
17      **else**
18         renF = BPM($v$, $w$)
19      df($v$, $w$) = min{insF, delF, renF}
20      renT = df($v$, $w$) + $\gamma'(v, w)$
21      dt($v$, $w$) = min{insT, delT, renT}
22  **return** $dt(root(T_1), root(T_2))$

---

case and a sequence edit distance computation in the ordered case. Since the sequence edit distance cannot be smaller than the bipartite graph matching cost, we focus on the bipartite graph matching.

Figure 5 illustrates the bipartite graph for the nodes $chd(v)$ and $chd(w)$. The edge cost $cost(c_i, c'_j)$ between two nodes $c_i \in chd(v)$ and $c'_j \in chd(w)$ is the tree distance between the subtrees rooted in these nodes, $dt(c_i, c'_j)$. To simplify the presentation, we assume $l = deg(v) < deg(w) = m$, i.e., $k = m - l$ subtrees will be matched to the empty tree. We denote the cost of the bipartite matching between the children of two nodes $v, w$ with $BPM(v, w)$.

*Aggregate Size Bound.* To establish a lower bound on the bipartite graph matching cost, we leverage the specific characteristics of the edge costs in our scenario. Since the edge costs are given by the respective subtree distances, we can bound the cost by the size difference of the subtrees, $cost^*(c_i, c'_j) = |(|T[c_i]| - |T[c'_j]|)| \leq cost(c_i, c'_j)$. A minimal matching $BPM^*(v, w)$ that uses $cost^*(c_i, c'_j)$ cannot be more expensive than the original matching, $BPM^*(v, w) \leq BPM(v, w)$. We leverage this fact to derive a novel lower bound based on subtree sizes. We define the *sorted aggregate size* between start $s$ and end $e$ in a subforest $F[v]$ as

$$SAS(v, s, e) = \sum_{i=s}^{e} |T[c_i]|, \quad c_i \in chd(v), \quad (6)$$

where $c_i$ is the $i$-th smallest subtree in $F[v]$ (ties broken arbitrarily).

The intuition of our bound is as follows: There exists a matching with cost $BPM(v, w)$, $deg(v) < deg(w)$, that matches the $k = deg(w) - deg(v)$ smallest subtrees to the empty tree, inducing cost $SAS(w, 1, k)$. The matching cost between the remaining subtrees is no larger than the difference of their aggregate subtree sizes.

THEOREM 2 (AGGREGATE SIZE BOUND). *Given two JSON tree nodes $v \in T_1, w \in T_2$. Let $d_v = deg(v)$, $d_w = deg(w)$, $k = d_w - d_v$, and $d_v \leq d_w$, then:*

$$BPM(v, w) \geq |SAS(v, 1, d_v) - SAS(w, k+1, d_w)| + SAS(w, 1, k).$$

PROOF. We show that the lower bound holds for $BPM^*(v, w) \leq BPM(v, w)$ with edge costs $cost^*(c_i, c'_j) = ||T[c_i]| - |T[c'_j]||$, $c_i \in chd(v)$, and $c'_j \in chd(w)$. W.l.o.g., assume that the children of $w$ are sorted by subtree size, $|T[c'_i]| \leq |T[c'_j]|$ if $i < j$. (a) We show that there is a min-cost matching that maps the $k$ smallest subtrees to the empty tree, inducing cost $|SAS(w, 1, k)|$: Let $\mathcal{M} \subseteq chd(v) \times chd(w)$ be a min-cost matching with $(\epsilon, c'_j) \in \mathcal{M}$ such that $j > k$, i.e., $|T[c'_j]| \geq |T[c'_k]|$. Then, there must be a node $c'_x$ with $x \leq k$ that is mapped to node $c_i$, $c_i \neq \epsilon$. By changing the matching to $\mathcal{M} \setminus \{(\epsilon, c'_j), (c_i, c'_x)\} \cup \{(c_i, c'_j), (\epsilon, c'_x)\}$, the cost cannot increase as can be verified by considering the three options for the size of $T[c_i]$: $|T[c'_j]| \leq |T[c_i]|$ (cost decreases by $2(|T[c'_j]| - |T[c'_x]|)$), $|T[c'_x]| \leq |T[c_i]| < |T[c'_j]|$ (cost decreases by $2(|T[c_i]| - |T[c'_x]|)$), and $|T[c_i]| < |T[c'_x]|$ (cost does not change). We repeat this step until all nodes $c'_1, c'_2, \ldots, c'_k$ are mapped to empty nodes. (b) The cost of the remaining matching $\mathcal{M}' = \{(c_i, c'_j) \in \mathcal{M} \mid c_i \neq \epsilon\}$ is $\sum_{(c_i, c'_j) \in \mathcal{M}'} |(|T[c_i]| - |T[c'_j]|)|$, which is at least $|SAS(v, 1, d_v) - SAS(w, k+1, d_w)| = \sum_{(c_i, c'_j) \in \mathcal{M}'} (|T[c_i]| - |T[c'_j]|)$. □

EXAMPLE 5. *For the root nodes of the JSON trees in Figure 2, $BMP(v_{11}, w_9) = 5$ and the aggregate size bound is 2 ($SAS(w_9, 1, k) = 0$ since both nodes have the same degree, i.e., $k = 0$). In Figure 6, $k = 1$ and the aggregate subtree bound is 9: $SAS(w, 1, k) = 5$ and $|SAS(v, 1, d_v) - SAS(w, k+1, d_w)| = 4$. Note that our aggregate size bound performs much better than a simple subtree size difference bound, which is $|SAS(v, 1, d_v) - SAS(w, 1, d_w)| = 1$ in this example.*

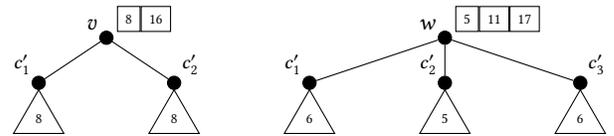

**Figure 6:** *SAS* arrays for the aggregate size bound.

*Efficient Computation of Aggregate Size Bound.* The aggregate size bound requires us to compute sums of subtree sizes. Since the bound is computed $O(|T_1||T_2|)$ times (for all pairs of parent nodes), computing these sums is too expensive. We precompute an array $SAS_v$ of size $deg(v)$ for each node $v \in T_1$ with $SAS_v[i] = SAS(v, 1, i)$ (cf. Eq. (6)); analogously $SAS_w$ for all $w \in T_2$ is computed. Thanks to the SAS arrays we can compute the bound in constant time:

$$|SAS(v, 1, d_v) - SAS(w, k+1, d_w)| + SAS(w, 1, k) =$$
$$|SAS_v[d_v] - SAS_w[d_w] + SAS_w[k]| + SAS_w[k] \quad (7)$$

EXAMPLE 6. $SAS_{w_9} = [2, 4, 8]$ *for root node* $w_9$ *in Figure 2. Figure 6 shows the* SAS *arrays for the root nodes* $v$ *and* $w$ *of the example trees.*

*Local Greedy Lower Bound.* The *local greedy lower bound* on $BPM(v, w)$ matches each node by following the lowest cost edge. The result may violate the one-to-one requirement and therefore may not be a valid matching. Similar bounds have been used before (e.g., [45]). Since this bound is as expensive as the sequence edit distance (quadratic in the node degrees as all edge costs must be checked), it is only useful for the bipartite graph matching.

LEMMA 2 (LOCAL GREEDY LOWER BOUND). *Let* $T_1, T_2$ *be JSON trees,* $v \in T_1, w \in T_2$. *Let* $GM_v \subseteq chd(v) \times chd(w)$ *map* $c_i \in chd(v)$ *to some* $c_j \in chd(w)$ *such that* $dt(c_i, c_j)$ *is minimal;* $GM_w \subseteq chd(w) \times chd(v)$ *is defined analogously:*

$$BPM(v, w) \geq \max\{\gamma(GM_v), \gamma(GM_w)\}$$

PROOF. $GM_v$ greedily picks the lowest cost outgoing edge for each $c_i \in chd(v)$. The bipartite matching must also provide a one-to-one matching and may therefore not be able to use the lowest cost edge for each node. Analogous reasoning for $GM_w$. □

We show how to compute $GM_v$ and $GM_w$ with low overhead: While we build the bipartite graph and retrieve all edge costs between the children of two nodes $v, w$, we maintain the minimum cost edge for each node $c_i \in chd(v)$ and $c'_j \in chd(w)$. In a single pass over the nodes, we get $GM_v$ and $GM_w$ with linear overhead.

An interesting opportunity arises when $GM_v$ or $GM_w$ is one-to-one: In this case, we can skip the bipartite graph matching since $BPM(v, w) = \max\{\gamma(GM_v), \gamma(GM_w)\}$ and we know the exact costs.

## 3.3 The QuickJEDI Algorithm

We present QuickJEDI, our efficient algorithm for computing the JSON edit distance. QuickJEDI extends JEDI-baseline (Algorithm 1) with the results in Section 3.2. While the baseline must compute the expensive min-cost matching between the children of each node pair $(v, w)$, QuickJEDI checks the aggregate size bound (cf. Th. 2) to assess whether the matching is required. The aggregate size bound is a lower bound for both types of min-cost matchings: the sequence edit distance, $SED(v, w)$, for pairs of array nodes, and the bipartite graph matching, $BPM(v, w)$, which is applied otherwise. Only if the lower bound is smaller than both $insF$ and $delF$ (line 2), the min-cost matching must be computed. Before computing $BPM(v, w)$, we also check the local greedy lower bound (cf. Lemma 2).

We further avoid the min-cost matching for two special cases (omitted in Algorithm 2 for brevity): if both $v$ and $w$ are key nodes, they have only one child each ($c_v$ resp. $c_w$), and $renF = dt(c_v, c_w)$. If both $v$ and $w$ are literal values, they are leaves, and $renF = 0$.

## 4 THE JEDIORDER FILTER

In this section, we propose *JediOrder*, a highly effective upper bound filter on the JSON edit distance. In a JSON similarity query, the upper bound is evaluated before JEDI: if the upper bound is within the similarity threshold $\tau$, the expensive JEDI needs not be computed.

---

**Algorithm 2:** QuickJEDI($T_1, T_2$)

**Input:** JSON trees $T_1$ and $T_2$.
**Result:** JSON Edit Distance: JEDI($T_1, T_2$).

/* Lines 1-14 from Algorithm 1                                 */
1   $AggSizeBd = |SAS_v[d_v] - SAS_w[d_w] + SAS_w[k]| + SAS_w[k]$
2   **if** $AggSizeBd < \min\{insF, delF\}$ **then**
3     **if** $type(v) == type(w) ==$ array **then**
4       $renF = SED(v, w)$
5     **else**
6       $LocalGreedyBd = \max\{\gamma(GM_v), \gamma(GM_w)\}$
7       **if** $LocalGreedyBd < \min\{insF, delF\}$ **then**
8         $renF = BPM(v, w)$
/* Lines 19-21 from Algorithm 1                                */
9   **return** $dt(root(T_1), root(T_2))$

---

We discuss Wang's algorithm [55], the fastest known algorithm that (with some adaptions to JSON trees) computes JediOrder. Wang's algorithm is faster than JEDI (quadratic vs. cubic) and requires less space. It turns out, however, that Wang's algorithm is still too slow to be used as an upper bound filter. The upper bound is computed for all tree pairs, but can only avoid the JEDI computation when the upper bound is within the threshold $\tau$. Whenever the upper bound is larger than $\tau$ (including the cases when the true distance is larger), JEDI must be computed *in addition* to JediOrder.

To pay off, the upper bound filter must incur very low cost compared to the computation of the exact distance. To this end, we develop a new algorithm, called *JOFilter*, that takes the similarity threshold $\tau$ into account. JOFilter only assesses whether JediOrder is within threshold $\tau$ (which is enough for the filter purpose) and avoids computing the exact JediOrder value otherwise. With a clever tree traversal that considers only relevant node pairs, we achieve linear runtime (vs. quadratic runtime of Wang's algorithm).

## 4.1 Tree Sorting and Upper Bound Guarantee

JediOrder sorts the children of object nodes in a JSON document lexicographically by their keys; recall that the keys are string literals that are unique within an object. The result is an *ordered* JSON tree in which all sibling collections are totally ordered (cf. Figure 7). JediOrder computes the minimal, edit-based distance between sorted JSON trees. Thanks to the order, JediOrder does not need to compute a bipartite graph matching, $BPM(v, w)$, between the children of two nodes $v$ and $w$; instead, the cheaper sequence edit distance, $SED(v, w)$, is evaluated (cf. Section 3.1). Formally, JediOrder is defined as the cost of the min-cost mapping that satisfies Definition 2.

DEFINITION 2 (ORDERED JSON EDIT MAPPING). *A JSON edit mapping* $M$ *is ordered iff for any node pairs* $(v, w), (v', w') \in M$:

• $v$ *is to the left*[1] *of* $v'$ *iff* $w$ *is to the left of* $w'$ [order].

EXAMPLE 7. *Considering the (ordered) JSON edit mappings in Figures 2 and 7,* $JEDI(T_1, T_2) = 5$ *vs.* $JediOrder(T_1, T_2) = 8$. *Due to the lexicographical order of the key nodes in Figure 7, the node pairs* $(v_9, w_5)$ *and* $(v_{10}, w_6)$ *violate the order constraint and are not in the minimum-cost ordered JSON edit mapping.*

---

[1] $v$ is to the left of $v'$ if $v$ is not a descendant of $v'$ and precedes $v'$ in postorder.

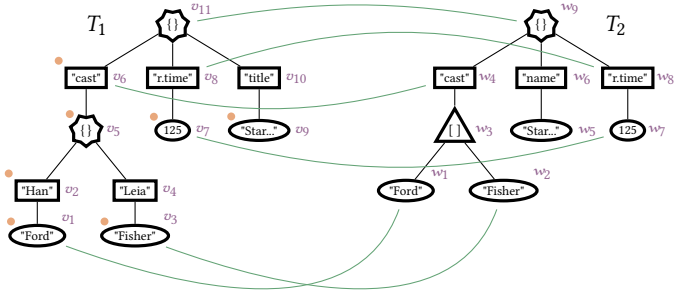

**Figure 7: Sorted JSON trees from Figure 2 including the ordered JSON edit mapping –, postorder numbers $v_i$ and $w_j$, and favorable children •.**

The *order* constraint in Definition 2 subsumes the *array-order* in Definition 1, thus JediOrder provides an upper bound for JEDI.

THEOREM 3 (JEDIORDER UPPER BOUND). *Given JSON trees $T_1$, $T_2$, then $JediOrder(T_1, T_2) \geq JEDI(T_1, T_2)$.*

PROOF. By contradiction. Assume an ordered JSON edit mapping $M_{JO}$ with cost $\gamma(M_{JO}) = JediOrder(T_1, T_2)$ and a JSON mapping $M$ with cost $\gamma(M) = JEDI(T_1, T_2)$ such that $JediOrder(T_1, T_2) < JEDI(T_1, T_2)$. Since $M_{JO}$ by Def. 2 is also a JSON edit mapping, $\gamma(M)$ is not minimal, which contradicts $\gamma(M) = JEDI(T_1, T_2)$. □

### 4.2 JediOrder Baseline: Wang's Algorithm

JediOrder is based on sorted, hence, ordered trees. As a baseline algorithm for JediOrder, we adapt the state-of-the-art constraint tree edit distance algorithm by Wang and Zhang [55], which runs in $O(|T_1||T_2|)$ time and $O(|T_2| \log |T_1|)$ space, to JSON trees.

*Recursive Solution:* The recursive solution discussed in Section 3.1 (cf. Eq. 1-3 and Lemma 1) also holds for JediOrder. Due to the total order among siblings, the minimum-cost matching in Eq. 2c is always computed by the sequence edit distance (rather than the more expensive bipartite graph matching). Zhang [60] shows the correctness of the recursion.

*Memory Efficient Implementation:* Similar to Algorithm 1, Wang's algorithm uses dynamic programming and a nested loop over all node pairs of the input trees $T_1$ and $T_2$. To reduce the memory complexity, Wang implements two key ideas: (1) The deletion and rename costs of a node $v \in T_1$ (delF/delT and renF/renT in Algorithm 1) w.r.t. all nodes $w \in T_2$ (inner loop) are computed *incrementally* while the children of $v$ are processed (in the outer loop). The required cost arrays of size $|T_2|$ are maintained with each node $v$; they are allocated when the first child of $v$ is processed and are released after processing $v$. (2) The nodes of $T_1$ (outer loop) are processed in *favorable child order*, a postorder traversal that visits the so-called favorable child (defined as the child with the largest subtree) first and all other children in the usual left-to-right order. This traversal guarantees that only $\log |T_1|$ nodes $v \in T_1$ maintain their cost arrays concurrently, thus reducing the memory complexity from quadratic to $O(|T_2| \log |T_1|)$. In Figure 7, the favorable children of $T_1$ are marked with an orange bullet •.

We will reuse these concepts and in addition leverage the similarity threshold to evaluate the JediOrder filter in linear time.

### 4.3 Leveraging the Distance Threshold

In the similarity lookup scenario, we are only interested in assessing whether JediOrder is within the similarity threshold $\tau$. Hence, we do not need to consider mappings $M_{JO}$ with a cost larger than $\tau$.

On top of the two optimizations of Wang's algorithm (cf. Section 4.2), we add a third key idea: (3) leverage the user-defined similarity threshold $\tau$ in combination with the postorder lower bound (cf. Lemma 3) to reduce the number of *relevant node pairs*.

LEMMA 3 (POSTORDER LOWER BOUND [31]). *Given an ordered JSON edit mapping $M_{JO}$ with cost $\gamma(M_{JO})$, for every node pair $(v, w) \in M_{JO}$ the following holds: $|post(v) - post(w)| \leq \gamma(M_{JO})$.*

PROOF. Proof by contradiction adapted from Hütter et al. [31]. Assume a node pair $(v, w) \in M_{JO}$ with $|post(v) - post(w)| > \gamma(M_{JO})$. Due to the ancestor (cf. Definition 1) and order (cf. Definition 2) constraints, the nodes $v_i$ with $post(v_i) < post(v)$ can only be mapped to the nodes $w_j$ with $post(w_j) < post(w)$. Since $|post(v) - post(w)| > \gamma(M_{JO})$, there must be at least $|post(v) - post(w)|$ unmapped nodes. Therefore, the overall mapping cost is greater than $\gamma(M_{JO})$ which contradicts the assumption. □

In similarity queries, the distance is bounded by the threshold $\tau$. Therefore, Lemma 3 implies that there are only $2\tau + 1$ eligible mapping partners $w \in T_2$ for a given node $v \in T_1$ such that the cost of the overall ordered JSON edit mapping is within $\tau$. We refer to the eligible nodes $w \in T_2$ as the $\tau$-*range* of a node $v \in T_1$.

EXAMPLE 8. *Consider the JSON trees in Figure 7 and a threshold $\tau = 2$. Any ordered JSON edit mapping that maps $v_6$ to a node in $T_2$ and has a cost of at most $\tau$ must map node $v_6$ to a node in its $\tau$-range, i.e., $w_4, w_5, w_6, w_7$, or $w_8$.*

Our goal is to apply the $\tau$-range in Wang's algorithm to avoid the nested loop over all node pairs. In particular, we strive to replace the inner loop over all nodes of $T_2$ by a constant $\tau$-range of $2\tau + 1$ nodes. This has an impact on the computation of the tree, the forest, and the sequence edit distance (SED) matrices.

In the tree and forest distance matrix, at most $2\tau + 1$ cells are filled per row. The other cells are guaranteed to exceed the threshold due to the $\tau$-range and do not need to be computed. Whenever these cells appear in a minimum computation, their value is considered to be infinite. If the overall mapping cost is within the threshold, the matrices store the correct JediOrder values. The correctness proof for the tree and the forest distance matrix is similar to the proof for the SED matrix, which we discuss in detail below.

We leverage the $\tau$-range also for SED, which is used to compute the minimum-cost matching between the ordered children of two nodes. A sequence edit matching must satisfy Definition 3.

DEFINITION 3 (SEQUENCE EDIT MATCHING). *Matching $M_{SED(m,n)} \subseteq chd(m) \times chd(n)$, $m \in T_1$ and $n \in T_2$, is a sequence edit matching iff for any pairs $(v, w), (v', w') \in M_{SED(m,n)}$ the following holds:*

- $v = v'$ iff $w = w'$ [one-to-one],
- $v$ *is to the left of* $v'$ *iff* $w$ *is to the left of* $w'$ [order].

Restricting SED to the $\tau$-range results in $\tau$-*restricted* SED matchings and the corresponding $\tau$-*sequence edit distance* ($\tau$SED).

JEDI: These aren't the JSON documents you're looking for... (Extended Version*)

DEFINITION 4 ($\tau$-RESTRICTED). *Let $M_{\tau SED(m,n)}$, $m \in T_1$ and $n \in T_2$, be a sequence edit matching. $M_{\tau SED(m,n)}$ is $\tau$-restricted iff for any pair $(v, w) \in M_{\tau SED(m,n)}$ the following holds:*

- $|post(v) - post(w)| \leq \tau$ [$\tau$-range].

The cost of a minimal SED matching $\gamma(M_{SED(m,n)})$ is identical to the cost of a minimal $\tau$-restricted SED matching $\gamma(M_{\tau SED(m,n)})$ whenever the overall JediOrder value is within the threshold $\tau$ (cf. Theorem 4). Otherwise, $\gamma(M_{\tau SED(m,n)})$ provides an upper bound on $\gamma(M_{SED(m,n)})$ and hence an upper bound on $\gamma(M_{JO})$ is computed. However, only tree pairs with $\gamma(M_{JO}) \leq \tau$ have to be considered in a similarity lookup.

THEOREM 4 (EXACT $\tau SED$). *If the minimal ordered JSON edit mapping $M_{JO}$ between $T_1$ and $T_2$ has a cost of $\gamma(M_{JO}) \leq \tau$, then $\gamma(M_{\tau SED(m,n)}) = \gamma(M_{SED(m,n)})$ for any node pair $(m, n) \in M_{JO}$.*

PROOF. By contradiction and case distinction. Suppose $\gamma(M_{JO}) \leq \tau$ and $\gamma(M_{\tau SED(m,n)}) \neq \gamma(M_{SED(m,n)})$. We consider the following cases:
*Case 1 ($\gamma(M_{\tau SED(m,n)}) < \gamma(M_{SED(m,n)})$):* A $\tau$-restricted sequence edit matching is also a sequence edit matching (cf. Definition 4) which contradicts the minimality of the sequence edit matching.
*Case 2.1 ($\gamma(M_{\tau SED(m,n)}) > \gamma(M_{SED(m,n)})$ and $\gamma(M_{JO}) = \tau$):* The $\tau$-restricted and unrestricted sequence edit matchings are identical by definition which contradicts $\gamma(M_{\tau SED(m,n)}) > \gamma(M_{SED(m,n)})$.
*Case 2.2 ($\gamma(M_{\tau SED(m,n)}) > \gamma(M_{SED(m,n)})$ and $\gamma(M_{JO}) < \tau$):* $|post(m) - post(n)| \leq \gamma(M_{JO})$ is stricter than $|post(m) - post(n)| \leq \tau$ which contradicts $\gamma(M_{\tau SED(m,n)}) > \gamma(M_{SED(m,n)})$. □

Note that $\tau SED$ is superior to a simple approach that uses a threshold on the string edit distance [44]. While $\tau SED$ prunes based on the postorder positions in the tree, the latter approach prunes based on the position in the string/sequence. Hence, for subtrees of size larger than one, $\tau SED$ provides better pruning power than the simple approach, and the same pruning power otherwise.

EXAMPLE 9. *Table 2 shows the SED matrix for the root nodes $v_{11}$ and $w_9$ of the trees in Figure 7. Consider node $v_8$ (at sequence position 2 and postorder 8) and a threshold $\tau = 2$. The unrestricted SED must compute all cells of the matrix. The simple threshold-based approach for the string edit distance must compute all cells for nodes with sequence positions $2 \pm 2$, i.e., all nodes $w_c \in chd(w_9)$ must be considered. $\tau SED$, however, only computes the cells in the $\tau$-range of the postorder positions (highlighted in green), e.g., for node $v_8$ only nodes with postorder positions $8 \pm 2$ ($w_6$ and $w_8$) need to be considered.*

|  | | chd($w_9$) | | |
|---|---|---|---|---|
| **SED** | $\epsilon$ | $w_4$ | $w_6$ | $w_8$ |
| $\epsilon$ | 0 | 4 | 6 | 8 |
| $v_6$ | 6 | 4 | 6 | 8 |
| $v_8$ | 8 | 6 | 6 | 6 |
| $v_{10}$ | 10 | 8 | 7 | 8 |

(row labels: chd($v_{11}$))

**Table 2: SED($v_{11}$, $w_9$) matrix of the root nodes in Figure 7. For threshold $\tau$ = 2, $\tau SED$ only computes the cells highlighted in green.**

**Algorithm 3:** Wang($T_1, T_2, \tau$)
**Input:** JSON trees $T_1$ and $T_2$, and threshold $\tau$.
/* Outline of SED computation in Wang's algorithm. */
1 **for** $v$ **in** $T_1$ **do**
2     **for** $w$ **in** $T_2$ **with** $|post(v) - post(w)| \leq \tau$ **do**
3        **for** $c$ **in** $chd(w)$ **do**
4           Compute cell (v,c) of the SED(p(v), w) matrix.

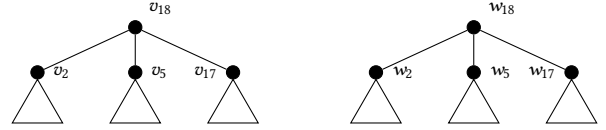

**Figure 8: Two identical JSON trees $T_1$ and $T_2$.**

## 4.4 Challenges of Applying the $\tau$-Range

For the loop variables $v \in T_1$ and $w \in T_2$, Wang computes row $v$ of the SED($p(v), w$) matrix (cf. Algorithm 3). This matrix has a row for each child of $p(v)$ and a column for each child of $w$.

To apply the $\tau$-range in Wang's algorithm, (1) the inner loop over all node pairs must be restricted to the nodes in the $\tau$-range and (2) the SED must be $\tau$-restricted. Unfortunately, extending Wang's algorithm with the $\tau$-range (highlighted in line 2, Algorithm 3) will lead to incorrect results. Consider the matrix of the SED($v_{18}, w_{18}$) computation between the two identical JSON trees in Figure 8 with $\tau = 2$. Wang's algorithm computes row $v_2$ while processing $v_2$ in the outer loop and $w_{18}$ in the inner loop. However, $w_{18}$ is not in the $\tau$-range of $v_2$; hence the node pair $(v_2, w_{18})$ is not considered in the nested loop, and row $v_2$ in the SED matrix is not filled.

## 4.5 The JOFilter Algorithm

We now present a novel algorithm, called *JOFilter*, that assesses whether JediOrder is within a given threshold $\tau$. Our solution is able to leverage all key ideas of the space-efficient algorithm by Wang (incremental cost computation and favorable child order, cf. Section 4.2) and the $\tau$-range introduced in Section 4.3. In the following, we discuss the key challenges that must be addressed and show that JOFilter runs in $O(n\tau)$ time and $O(n + \tau \log n)$ space.

*Cost arrays of size $\tau$.* Similar to Wang's algorithm, we split the auxiliary matrices into rows and store each row with the relevant nodes $v \in T_1$. A node $v$ maintains the following data: row $v$ of (1) the tree and (2) the forest distance matrix, denoted $v.dt$ and $v.df$, respectively; (3) the tree distance matrix row of $v$'s favorable child, $v.dt_{fc}$; finally, (4) two rows of the SED($v, w$) matrices for all $w \in T_2$, denoted $v.sed_{L_0}$ and $v.sed_{L_1}$, which are sufficient to compute SED [28]. Due to the $\tau$-range, the size of these cost arrays (i.e., matrix rows) can be reduced from $O(|T_2|)$ in Wang's algorithm to $O(\tau)$ in JOFilter. Summarizing, a node $v \in T_1$ stores auxiliary data of size $O(\tau)$. Moreover, the insertion (resp. deletion) costs of node $v$ in the forest, tree, and SED matrices, denoted $v.df_\epsilon$, $v.dt_\epsilon$, and $v.sed_\epsilon$, are stored in global arrays of size $|T_2|$.

*Logarithmic number of active nodes.* A node is called *active* while the node and its auxiliary data are held in main memory. A node $v$

**Algorithm 4:** JOFilter($T_1, T_2, \tau$)

**Input:** JSON trees $T_1$ and $T_2$, and threshold $\tau$.
/* Outline of SED computation in the JOFilter. */
1 **for** $v$ **in** $T_1$ **do**
2     **for** $w$ **in** $T_2$ **with** $|post(v) - post(w)| \leq \tau$ **do**
3         Compute cell (v,w) of the SED(p(v), p(w)) matrix.

becomes active when its favorable child is processed and inactive after $v$ itself was processed. The favorable child order guarantees that at most $O(\log |T_1|)$ nodes are active at any time [55].

*Applying the $\tau$-range.* We apply the $\tau$-range by replacing the inner loop over all nodes $w \in T_2$ by a constant range of $2\tau + 1$ nodes. As shown in Section 4.4, applying the $\tau$-range in Wang's algorithm leads to incorrect results. We therefore adapt the computation order of the values in the SED computation as shown in Algorithm 4: only a single cell $(v, w)$ of the SED$(p(v), p(w))$ matrix is filled in the inner loop rather than an entire matrix row. Since $v$ and $w$ are the loop variables, we guarantee that all node pairs in the $\tau$-range are considered in the SED computation.

*Algorithm.* We present the pseudocode of our solution, JOFilter, in Algorithm 5. Note that $ls(v)$ denotes the left sibling of node $v$ and a dot '.' accesses the data of a given node. The nodes $v \in T_1$ (outer loop) are traversed in favorable child order, while the nodes $w \in T_2$ (inner loop) are traversed in postorder. To avoid the computation between all node pairs, we apply the $\tau$-range (cf. Lemma 3) in the inner loop. Assuming that the rename and deletion costs are given, we first compute the tree and forest distance between nodes $v$ and $w$ (cf. lines 4-9). In the remainder of the algorithm (cf. lines 10-28), the deletion and rename costs for the parent of node $v$ are computed incrementally. After processing all node pairs, the overall distance is stored in the tree distance matrix line of the root node of $T_1$, $root(T_1).dt(root(T_2))$. The filter only accepts a tree pair $(T_1, T_2)$ iff JediOrder$(T_1, T_2) \leq \tau$ (cf. line 29).

*Complexity.* For each node $v \in T_1$ only $2\tau + 1$ nodes $w \in T_2$ are considered (cf. Lemma 3). Therefore, the overall *time* complexity is $O(|T_1|\tau)$. The space complexity is dominated by the global arrays of size $O(|T_2|)$ that store the insertion (resp. deletion) costs of node $v$ in the forest, tree, and SED matrices. Each active node fits in $O(\tau)$ space and there are at most $O(\log |T_1|)$ active nodes at any point in time, leading to an overall space complexity of $O(|T_2| + \tau \log |T_1|)$.

## 5 JSIM: JSON SIMILARITY INDEX

We now present the *JSIM* index for JSON similarity queries and discuss the use of index and filters in the similarity query context.

The input to the JSIM index over a tree database $\mathcal{T}$ is a query tree $T_q$ and a threshold $\tau$, the output is a *candidate set* $C \subseteq \mathcal{T}$ that is a superset of the query result, $R = \{T_i \in \mathcal{T} \mid JEDI(T_q, T_i) \leq \tau\} \subseteq C$.

Existing indexing techniques for tree similarity queries [31, 49] require ordered trees. They leverage concepts that (due to the missing order of object nodes) are not applicable to JSON, e.g., the postorder position of nodes in the tree [31] or an order-based partitioning of trees into subgraphs [49]. Sorting JSON trees does not solve the problem: The distance between sorted trees may increase w.r.t. JEDI such that the index fails to retrieve relevant trees.

**Algorithm 5:** JOFilter($T_1, T_2, \tau$)

**Input:** JSON trees $T_1$ and $T_2$, and threshold $\tau$.
**Result:** *True* if JediOrder$(T_1, T_2) \leq \tau$, *False* otherwise.
1 **for** $v$ **in** $T_1$ **do**     /* Favorable child order */
2     p = p(v)
3     **for** $w$ **with** $|post(v) - post(w)| \leq \tau$ **do** /* Postorder */
        /* Cost for inserting node $w$. */
4         insF = w.df$_\epsilon$ + min$_{c \in chd(w)}$\{v.df(c) - c.df$_\epsilon$\}
5         insT = w.dt$_\epsilon$ + min$_{c \in chd(w)}$\{v.dt(c) - c.dt$_\epsilon$\}
        /* Costs for deleting and renaming already computed. */
6         renF = v.sed$_{L_0}(w_t)$   /* $w$'s rightmost child $w_t$. */
7         v.df(w) = min\{insF, v.delF(w), renF\}
8         renT = v.df(w) + min\{$\gamma(v, w), \gamma(v, \lambda), \gamma(\lambda, w)$\}
9         v.dt(w) = min\{insT, v.delT(w), renT\}
        /* Compute deletion and rename costs for parent. */
10         **if** $v$ is favorable child **then**
11             p.dt$_{fc}$(w) = v.dt(w)
12             p.delF(w) = v.df$_\epsilon$ + v.df(w) - v.df$_\epsilon$
13             p.delT(w) = v.dt$_\epsilon$ + v.dt(w) - v.dt$_\epsilon$
14         **else**
15             p.delF(w) = min\{p.delF(w), p.df(0)+v.df(w)-v.df(0)\}
16             p.delT(w) = min\{p.delT(w), p.dt(0)+v.dt(w)-v.dt(0)\}
17         **if** $v$ is left-most child **then**
18             p.sed$_{L_1}$(0) = p.sed$_{L_0}$(0) + v.dt$_\epsilon$
19             p.sed$_{L_1}$(w) = min\{p.sed$_{L_1}$(ls(w)) + w.dt$_\epsilon$, w.sed$_\epsilon$ + v.dt$_\epsilon$, ls(w).sed$_\epsilon$ + v.dt(w)\}
20         **else if** $v$ is not favorable child **then**
21             p.sed$_{L_1}$(0) = p.sed$_{L_0}$(0) + v.dt$_\epsilon$
22             p.sed$_{L_1}$(w) = min\{p.sed$_{L_1}$(ls(w)) + w.dt$_\epsilon$, p.sed$_{L_0}$(w) + v.dt$_\epsilon$, p.sed$_{L_0}$(ls(w)) + v.dt(w)\}
23     **if** $v$ is left sibling of favorable child $c_f$ **then**
24         p.sed$_{L_0}$(0) = p.sed$_{L_1}$(0) + p.dt$_{fc}$(0)
25         **for** $w$ **with** $|post(v) - post(w)| \leq \tau$ **do**
26             p.sed$_{L_0}$(w) = min\{p.sed$_{L_0}$(ls(w)) + w.dt$_\epsilon$, p.sed$_{L_0}$(w) + p.dt$_{fc}$(0), p.sed$_{L_0}$(ls(w)) + p.dt$_{fc}$(w)\}
27     **else**
28         p.sed$_{L_0}$ = p.sed$_{L_1}$
29 **return** $root(T_1).dt(root(T_2)) \leq \tau$

Our JSIM index leverages a novel lower bound for JSON trees, called *JSON region bound*, that is based on the position of a node in the tree. Based on this lower bound and a node label filter, we build an effective multi-level index that only returns trees $T_i \in \mathcal{T}$ that pass all filters. Moreover, we introduce a technique that decreases the search threshold level by level during the index lookup. This allows us to aggressively prune index branches at deeper index levels.

### 5.1 Leveraging Node Position and Labels

We now present the JSON region bound that is based on the ancestor constraint of the JSON edit mapping in Definition 1. Assume that the node pair $(v, w)$ in Figure 9 is mapped; then, $anc(v)$ must be mapped to $anc(w)$ (red), $desc(v)$ to $desc(w)$ (green), and $lr(v)$ to $lr(w)$ (blue). The left-right nodes $lr(v)$ of node $v$ are all nodes in $T_q$ different from $v$, $desc(v)$, and $anc(v)$. Intuitively, the size difference of the individual regions imposes a lower bound on the respective mapping cost. For example, the cost of mapping the ancestors in Figure 9 is at least one.

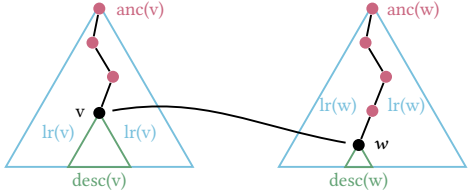

Figure 9: Due to the ancestor constraint, mapping node $v$ to $w$ splits the JSON tree into three regions.

LEMMA 4 (JSON REGION BOUND). *Let $T_1$, $T_2$ be JSON trees, $M$ a JSON edit mapping from $T_1$ to $T_2$. For a given similarity threshold $\tau$, if the cost of the mapping is $\gamma(M) \leq \tau$, then for each $(v, w) \in M$:*

$$||desc(v)| - |desc(w)|| + ||anc(v)| - |anc(w)|| + ||lr(v)| - |lr(w)|| \leq \tau.$$

PROOF. The proof is by contradiction: Suppose $\gamma(M_{DPJED}) \leq \tau$ and $||desc(v)| - |desc(w)|| + ||anc(v)| - |anc(w)|| + ||lr(v)| - |lr(w)|| > \tau$. Due to the ancestor constraint of the edit mapping (cf. Definition 1) any $(v', w') \in M_{DPJED}$ must either be in $desc(v) \times desc(w)$, $anc(v) \times anc(w)$, or $lr(v) \times lr(w)$. From the initial assumption, we know that more than $\tau$ edit operations are needed to adjust the sizes of these regions which contradicts $\gamma(M_{DPJED}) \leq \tau$. □

*Tightening the Bound.* An interesting observation is that when we know one of the size differences in Lemma 4, e.g., $\Delta = ||desc(v)| - |desc(w)||$, we can tighten the bound for the remaining, unknown differences: $||anc(v)| - |anc(w)|| + ||lr(v)| - |lr(w)|| \leq \tau - \Delta$. We leverage this effect to prune branches in our index traversal.

*Label Intersection.* A well known lower bound is based on the bag intersection of node labels [3]. For JSON trees, we need to replace node labels by (*label, type*) pairs. Then, the following holds:

$$JEDI(T_1, T_2) \geq max(|N(T_1)|, |N(T_2)|) - |N(T_1) \cap N(T_2)|. \quad (8)$$

## 5.2 Index Structure and Lookup

We discuss the structure of the JSIM index and our lookup technique that leverages the filters discussed in Section 5.1.

*Building the Index.* JSIM is a tree with four levels that store (1) node labels, (2) descendant counts, (3) ancestor counts, and (4) left-right node counts, respectively. Each index node is a sorted list of entries that either points to a child node (non-leaf entry) or to a list of indexed trees (leaf entries).

A new tree $T_i$ is inserted $|T_i|$ times into the index, once for each node. Each node adds a constant number of (at most 5) index entries. Therefore, the overall index size is proportional to the aggregated number of nodes of the indexed JSON trees. The insert path for a node $v \in T_i$ is determined by its label, its number of descendants, ancestors, and left-right nodes. New values are inserted into the respective index node, for existing values the child pointer is followed. The process of inserting node $v_8 \in T_1$ from Figure 2 into the index is highlighted in Figure 10a (green). Tree $T_1$ is inserted with label = "r.time", $|desc(v_8)| = 1$, $|anc(v_8)| = 1$, and $|lr(v_8)| = 8$.

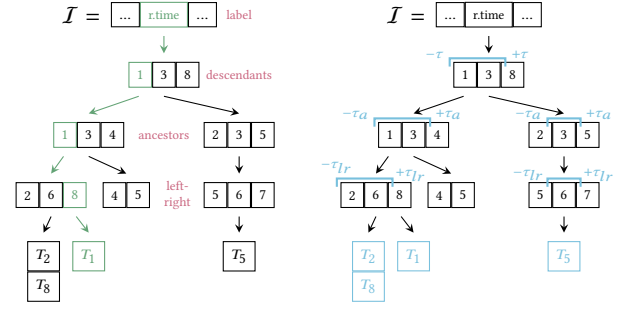

(a) Insert node $v_8$ from Figure 2. (b) Lookup node $w_8$ from Figure 2.

Figure 10: JSIM index: four-level inverted list.

*Index Lookup.* The lookup for query tree $T_q$ processes $\tau + 1$ nodes $v \in T_q$, and for each node proceeds in two steps: (1) *Label lookup:* Follow the branch for the label of $v$ in the index root node. The index lookup is limited to only $\tau + 1$ nodes since any tree $T_i$ that has more than $\tau + 1$ mismatching labels with $T_q$ cannot be within edit distance $\tau$ [31, 38]. (2) *Region traversal:* We leverage Lemma 4 to traverse the remaining levels. At each node, we follow all keys $k$ (i.e., region counts) that fall into the range given by Lemma 4, e.g., $d = ||desc(v)| - k| \leq \tau$ at the descendant count level. Note that each of the three size differences (which are all positive) must be within the threshold $\tau$. At the lower index levels, we leverage the size difference that we know from previous levels, e.g., the threshold for the ancestor level can be decreased to $\tau_a = \tau - d$ and the index verifies all keys $k_a$ with $a = ||anc(v)| - k_a| \leq \tau_a$. The process for the fourth level is similar, we verify all keys $k_{lr}$ with $||lr(v)| - k_{lr}| \leq \tau_{lr}$ against an even further reduced threshold $\tau_{lr} = \tau_a - a$. All trees in the lists that we reach are candidates and are returned by the index. For example, the lookup of node $w_8 \in T_2$ (Figure 2) is illustrated in Figure 10b (blue) and returns $T_2$, $T_8$, $T_1$, $T_5$.

Note that a search may end before reaching a leaf node when no trees in the $\tau$-range are found. This desirable effect is boosted by reducing the $\tau$-range at each level.

## 5.3 JSON Similarity Lookups

We leverage our techniques (i.e., JSIM, JOFilter, and QuickJEDI) to answer JSON similarity lookup queries as follows: (1) Lookup query tree $T_q$ with threshold $\tau$ in the JSIM index to retrieve candidate set $C$. (2) For each tree $T_i \in C$ check the label intersection lower bound in Eq. (8). (3) For the remaining candidates $T_i \in C'$, if $JediOrder(T_q, T_i) \leq \tau$, then $T_i$ is a result pair. (4) Verify the remaining candidates $T_i \in C''$ by computing $QuickJEDI(T_q, T_i)$.

## 6 EXPERIMENTS

We experimentally evaluate our solution for JSON similarity lookups on 22 real-world datasets in a unified C++ framework. The source code [30] and the experimental data [29] are publicly available. The experiments are executed single-threaded on an Intel Xeon E5-2630 v3 2.40GHz server with 16 cores and 96GB of RAM (Debian 10).

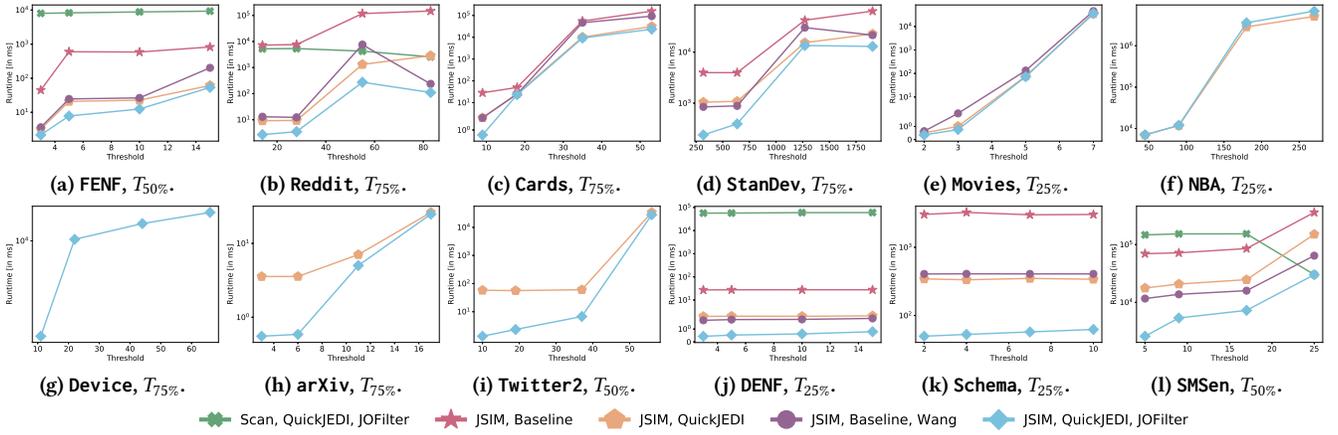

Figure 11: Overall runtime: JSON similarity lookup query.

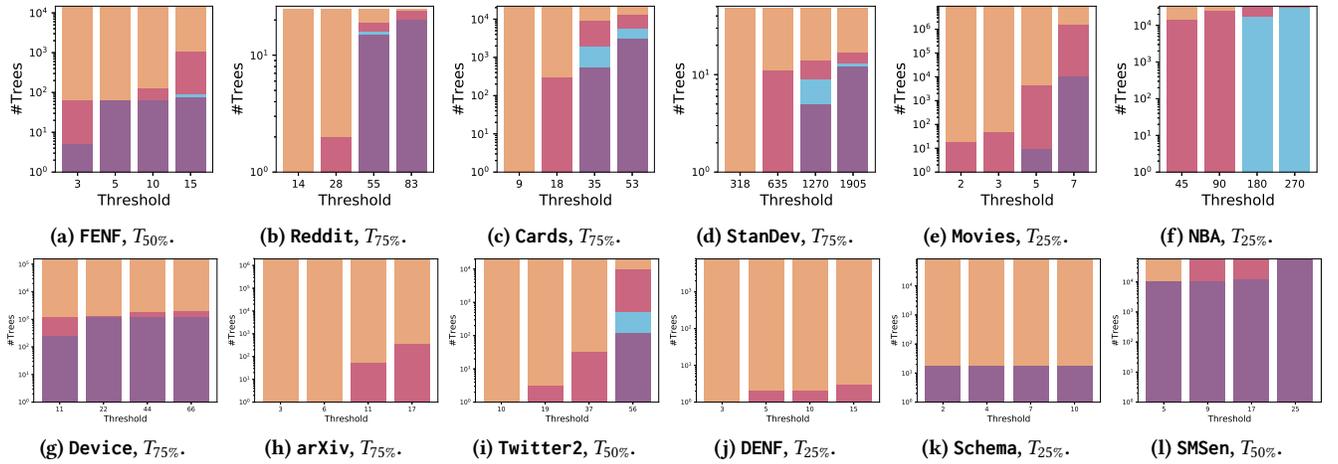

Figure 12: Filter effectiveness: pruned by the ■ JSIM index, ■ label intersection, ■ upper bound, and ■ number of verifications.

### 6.1 Setup

*Algorithms:* We evaluate various algorithmic combinations. *Scan* denotes a linear scan, *JSIM* denotes our index (cf. Section 5), *Wang* is the state-of-the-art JediOrder algorithm, *JOFilter* is our JediOrder filter (Algorithm 5), *Baseline* refers to the JEDI-baseline (Algorithm 1) and *QuickJEDI* to our optimized version (Algorithm 2).

*Datasets:* The evaluation is performed on a collection of 22 real-world JSON datasets. We summarize their most important characteristics: *Collection sizes* of up to 8.76 million JSON trees; *JSON tree sizes* of up to 48k nodes; a *type distribution* within a JSON tree of up to 20% objects, 10% arrays, 49% keys, and 49% literals; the *degree of object nodes* is typically less than 20 with the exception of one dataset (104); the *degree of array nodes* is up to 1603 values; one dataset provides a *depth* of 50 (less than 14 for all other datasets).

We briefly describe the datasets used for the experiments in Figures 11 and 12. (1) FENF [23]: FDA enforcement actions, ~14k documents with an average of 49 nodes per document and a depth of 3. (2) Reddit [43]: 25 Reddit articles with an average of 265 nodes per document. This dataset provides the highest object degree of 104 children. (3) Cards [27]: ~20k Magic cards with an average number of 132 nodes per document. (4) StanDev [53]: question-answering dataset, 48 documents with up to ~18k nodes and an average of 5,379 nodes per document. (5) Movies[40]: TV and movie ratings, ~8.7 million documents with an average of 23 nodes per document. (6) NBA [15]: ~31k NBA games with an average of 977 nodes per document. (7) Device [23]: ~150k FDA enforcement actions with up to 3,264 nodes per document. (8) arXiv [52]: 1.8 million research publications with an average of 53 nodes per publication. (9) Twitter2 [51]: ~19k tweets with an average of 195 nodes per document. (10) DENF [23]: ~7k FDA enforcement actions with an average of 59 nodes per document. (11) Schema [5]: 81k JSON schemas with up to 48k nodes per schema document. (12) SMSen [14]: ~55k SMS messages with an average of 81 nodes per document.

*Experimental Setup:* For each dataset, we perform JSON similarity lookup queries for three different query trees and four different thresholds. Since the runtime of the distance algorithms depends on the tree sizes, we pick the query trees that are closest to the 25%, 50%, and 75% quantiles of the tree sizes for each dataset (denoted $T_{25\%}$, $T_{50\%}$, and $T_{75\%}$). The goal of similarity lookup queries is to return documents that are similar to the query document, hence useful thresholds depend on the size of the query tree. We experiment with thresholds that are 5%, 10%, 20%, and 30% of the respective query tree size. The timeout for computing the results for all thresholds for a given algorithm and dataset is 24 hours.

*Evaluation:* We analyze the overall runtime and the effectiveness of the introduced bounds. Each plot in Figure 11 and 12 shows the results of a single experiment, i.e., a given dataset and query tree for varying thresholds on the x-axis. For example, Figure 11a shows the results for dataset FENF and the 50% quantile query tree $T_{50\%}$.

Figure 11 shows the overall lookup runtimes in milliseconds for various algorithm combinations. Figure 12 evaluates the number of trees pruned by the individual filters as well as the number of required verifications. The total height of a bar is the number of documents (i.e., trees) in the dataset, the colors distinguish the tree pairs that are pruned by the JSIM index (orange), the label intersection (red), the upper bound (purple), and the number of verifications (blue). The runtime and effectiveness plots are aligned, e.g., Figures 11a and 12a result from the same experiment.

The overall experiment includes 66 dataset/query combinations. Due to space restrictions, we provide a representative selection that covers the most relevant phenomena (cf. Figures 11 and 12).

## 6.2 Results

*JSIM Index vs. Dataset Scan.* We measure the effectiveness of the index by the number of returned candidates (cf. Figure 12). Especially for small thresholds, the returned candidates are orders of magnitude smaller than the collection size (e.g., Figure 12h and 12e).

Due to the smaller number of candidates, the index outperforms the scan in each experiment, e.g., the index is up to five orders of magnitude faster in Figure 11j. For larger datasets (e.g., Figure 11e), the index is needed to answer the query within the timeout. In some scenarios, however, applying an index without further optimizations is not enough: The Reddit dataset used for the experiment in Figure 11b only contains 25 documents; however, due to its object degree of 104, verifying even a single candidate significantly increases the runtime. As a result, scanning Reddit with our optimized algorithms (Scan, QuickJEDI, JOFilter) outperforms the index-based solution with baseline verification (JSIM, Baseline). In some cases, when the lookup result is empty, the query is answered only within the index, i.e., neither the upper bound nor the verification are computed (e.g., Figures 12h and 12j).

*Wang vs. JOFilter.* Next, we compare the state-of-the-art JediOrder algorithm (Wang) with our optimized algorithm (JOFilter). The experimental results show the behaviour expected based on the runtime complexities of the algorithms. The complexity of JOFilter depends on the threshold. Even for larger thresholds, JOFilter is superior to Wang due to the quadratic complexity of the latter. We compare the runtimes of Wang (purple) and JOFilter (blue) in Figure 11. In Figure 11a, no candidates must be verified except for threshold 15; hence the runtime improves from JOFilter alone. We observe the largest improvements of JOFilter in Figures 11k and 11l, where Wang is up to an order of magnitude slower.

In many scenarios (cf. Figures 12), the upper bound identifies most of the result set and only few trees must be verified (blue bar). However, the upper bound is applied to each candidate and introduces additional overhead which may increase the runtime in cases where the upper bound is not effective (cf. Figure 11f). These results show that an efficient verification algorithm is indispensable.

*Baseline Verification vs. QuickJEDI.* We also evaluated the effect of the optimized verification algorithm QuickJEDI over the baseline without applying the upper bound (red stars vs. orange pentagons in Figure 11). The complexities of both algorithms heavily depend on the degrees of the trees. QuickJEDI aims at skipping the expensive min-cost matching computation, which substantially reduces the runtime. Consider the measurements for threshold 55 in Figures 11b and 12b: even though only 16 trees have to be verified, the runtime difference between the baseline and QuickJEDI is almost two orders of magnitude. This results from the characteristics of the Reddit dataset, where some documents feature up to 104 unordered key-value pairs per object. Moreover, in 4 out of the 22 datasets the lookup terminated within the timeout only in configurations that include QuickJEDI (e.g., Figures 11f and 11i).

*Summary.* Overall, the best performance results are achieved by combining the JSIM index, JOFilter, and QuickJEDI. This configuration provides the lowest runtimes for 61 out of our total of 66 experiments and is the only one that is able to process all datasets (e.g., Figure 11g) within 24 hours. Only in cases where the upper bound is ineffective (cf. Figure 11f), QuickJEDI without JOFilter is slightly better. These results are robust even when the characteristics of the datasets vary, e.g., large documents (cf. Figure 11d) and large collections (cf. Figure 11e). In 37 experiments, the query is answered without applying a verification algorithm, i.e., the candidates returned by the index are equivalent to the result and are verified by the upper bound, highlighting the filter effectiveness.

## 7 RELATED WORK

*JSON Tree Representations.* There exist multiple tree representations of JSON documents. Bourhis et al. [7] represent keys and the array order as edges and values as leaf nodes; the approach by Shukla et al. [46] is similar, but keys and the array order are inner nodes instead of edges. Similar to our approach, Klettke et al. [34] introduce three different types of nodes (object, array, property) in addition to the label. Spoth et al. [47] use a tree containing atomic values at the leaves and complex values in the inner nodes. These representations either discard the object and the array information or encode the information in the edges of a tree; both choices are unsuitable for node edit operations. Tree representations of XML data (e.g., by Augsten et al. [2]) cannot be applied in the context of JSON since XML siblings are considered to be ordered.

*JSON Similarity.* To the best of our knowledge, there is only one scientific work on JSON diffs. Cao et al. [12] present an algorithm that computes a JSON patch based on the edit operations defined in RFC6902 [9]. In an experimental study, a comparison to four open source solutions was performed. However, the runtime and space complexity of the presented algorithm was not discussed. Further, the resulting patch is not minimal and therefore unsuitable for similarity queries. Yahia et al. [56] proposed a YAML-based language for describing change-detection strategies on JSON data.

Diff algorithms do exist for other hierarchical data formats. Chawathe et al. [13] present an algorithm that computes minimal diffs for LaTeX and HTML documents. The following edit operations are considered: insert and delete leaf nodes, update the value of any node, and subtree moves. The XML diffs by Cobena et al. [17] consider insertions and deletions of subtrees, value updates of any node, and moves of a node or a part of a subtree. Both approaches operate on ordered trees and are therefore unsuitable for JSON.

*JSON Schema.* Most of the scientific work related to JSON deal with schema extraction. Schemas are used as dataset descriptions or to enable optimization techniques in database systems. Durner et al. [19] present a solution to extract multiple local schemas for a single dataset. The schemas are grouped based on the label sets of the keys in a document. Baazizi et al. [4] introduce a parametric and parallel schema inference algorithm. Klettke et al. [34] present a schema extraction algorithm to identify structural outliers based on structure identification graphs. While the goal of schema extraction is different from that of similarity queries, JEDI could be used to identify schemas for similar documents.

*Tree Edit Distance.* A well-known edit distance for hierarchical data is the tree edit distance (TED). The current best algorithm for ordered trees by Pawlik and Augsten [41] computes TED in cubic time using quadratic memory. Computing TED for unordered trees is NP-hard [62]. Further, TED was applied for different query types, e.g., similarity joins [31] and top-k similarity joins [36]. However, these techniques are not applicable for JSON since JSON trees consist of ordered as well as unordered children. In fact, we showed that a TED adaption for JSON results in an NP-hard problem.

Zhang introduced a constraint TED version which can be computed in time $O(n^2)$ for ordered [60] and $O(|T_1| \cdot |T_2| \cdot (deg(T_1) + deg(T_2)) \cdot log_2(deg(T_1) + deg(T_2)))$ for unordered trees [61]. Similar to TED, both algorithms are designed for either ordered or unordered trees. We combined both approaches to construct the baseline JEDI algorithm. As shown in our experimental evaluation, we introduce heuristics that decrease the runtime of JEDI often by orders of magnitude. The ordered constraint TED algorithm by Wang et al. [55] using $O(n \log n)$ memory was used as a baseline algorithm for JediOrder. We introduced a novel JediOrder algorithm that improves the complexity to be linear in time and space.

*Heuristics for the Unordered Tree Edit Distance.* Due to the computational complexity of the unordered TED, a number of heuristics have been presented. Augsten et al. [2] introduced an approximation based on tree decomposition, called windowed pq-grams, that splits a tree into a set of smaller elements which are then compared to the decomposition of another tree. They experimentally showed that windowed pq-grams outperform other tree decomposition algorithms (binary branches [57], path shingles [11], and valid subtrees [25]). Rather than introducing approximations, we defined an exact and minimal JEDI distance.

## 8 CONCLUSION AND FUTURE WORK

In this paper, we addressed the problem of JSON similarity lookup queries: Given a query document $T_q$ and a distance threshold $\tau$, retrieve all documents from a JSON database $\mathcal{T}$ that are within distance $\tau$ from the query. We proposed (a) a lossless tree representation for JSON, (b) JEDI, the first edit-based distance for JSON documents, (c) the efficient QuickJEDI algorithm for JEDI, (d) the JSIM index to efficiently retrieve candidate trees for JSON similarity queries, and (e) JediOrder, an effective upper bound on JEDI. In our experiments, we scaled JSON similarity lookup queries to databases with millions of documents and JSON trees with thousands of nodes.

In an ongoing effort, our solution is being integrated into Apache AsterixDB, an open-source big data management system that uses partitioned-parallel query processing and a JSON-like data format.

## ACKNOWLEDGMENTS
We thank Wail Alkowaileet, Daniel Kocher, Mateusz Pawlik, and Zhihui Yang for valuable discussions. This work was supported by the Austrian Marshall Plan Foundation, the Austrian Science Fund (FWF): P 29859 and P 34962, the European Research Council (ERC) under the European Union's Horizon 2020 research and innovation program, under grant agreement No. 695412, and the Czech Ministry of Education, Youth and Sports from the Czech Operational Programme Research, Development, and Education, under grant agreement No. CZ.02.1.01/0.0/0.0/15_003/0000421.

JEDI: These aren't the JSON documents you're looking for... (Extended Version*)

# A PROOFS

## A.1 NP-completeness of DNJEDI

Removing the document-preserving constraint from Definition 1 results in the document-neglecting JSON edit distance (DNJEDI). To show the computational complexity of DNJEDI, we consider the following decision problem:

DEFINITION 5. **DNJEDIx**.
Instance: *Two JSON trees $T_S$ and $T_C$ and a positive integer $x$.*
Question: *Is the JSON edit distance $DNJEDI(T_S, T_C) \le x$?*

Similar to the unordered tree edit distance [59], the proof is by reducing the exact cover by 3-sets (X3C) problem, c.f. Definition 6 to the *DNJEDIx* problem. The main differences compared to the unordered tree edit distance are the two additional JSON edit mapping constraints (cf. Definition 1). Further, the tree transformation used by Zhang would produce invalid JSON trees and therefore needs to be changed.

DEFINITION 6. **Exact cover by 3-sets (X3C).**
Instance: *A finite set $S = \{s_1, s_2, \ldots, s_m\}$, where $m = 3k$, and a collection $C = \{C_1, C_2, \ldots, C_n\}$, where $C_i = \{c_{i_1}, c_{i_2}, c_{i_3}\}$ and $c_{i_j} \in S$.*
Question: *Is there a subcollection $C' \subseteq C$ such that every element of $S$ occurs in exactly one member of $C'$?*

Given an instance of *X3C*, let $l_1, \ldots, l_{2n-k}, l_x, l_y, l_z, L \notin S$ be strings that are not elements of set $S$ and, w.l.o.g., each element $s_i \in S$ occurs only once in $S$. The polynomial time transformation of an *X3C* instance into a *DNJEDIx* instance (i.e., JSON trees $T_S$, $T_C$, and integer $i$) is shown in Figure 13. Tree $T_S$ is built based on set $S$; for each element $s_i \in S$ a $\lambda$-subtree is inserted ($3k$ many in total), and additionally $n - k$ $\mu$-subtrees are inserted. Tree $T_C$ is built based on collection $C$; for each element $C_i \in C$, a $\kappa$-subtree is inserted ($n$ many in total). Further, let $i = 4n - 2k$.

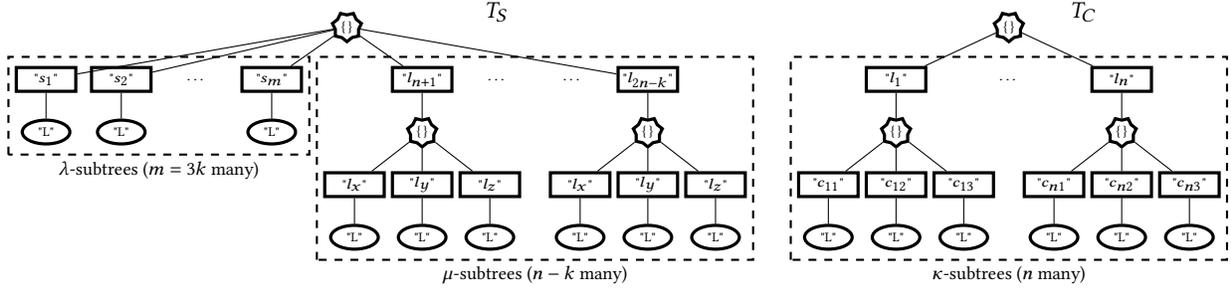

Figure 13: Polynomial time transformation of an exact cover by 3-sets (X3C) instance into JSON trees $T_S$ and $T_C$.

Given an instance of *X3C* with JSON tree transformations $T_C$ and $T_S$, we show that the following holds: $DNJEDI(T_S, T_C) \leq 4n - 2k$ (i.e., DNJEDIx with $i = 4n - 2k$) iff there is an exact cover by 3-sets.

LEMMA 5. *Given an instance of* X3C *with JSON tree transformations $T_S$ and $T_C$, let $M_{DNJ}$ be the minimal document-neglecting JSON edit mapping between JSON trees $T_S$ and $T_C$, then $DNJEDI(T_S, T_C) \geq 4n - 2k$.*

PROOF. As a result of the described transformation process, the sizes of the trees $|T_C|$, $|T_S|$ and the mapping size $|M_{DNJ}|$ are as follows:

- $|T_S| = 8(n-k) + 2(3k) + 1$: $n-k$ $\mu$-subtrees of size 8 are inserted, a $\lambda$-subtree with 2 nodes is inserted (key and literal) for each $s_i \in S$ with $|S| = 3k$, and one node for the root.
- $|T_C| = 8n + 1$: for each $C_i \in C$ with $|C| = n$, a $\kappa$-subtree with 8 nodes is inserted, and one node for the root.
- $|M_{DNJ}| = |T_S| - d = 8n - 2k + 1 - d$, where $d$ is the number of unmapped nodes in $T_S$. From the JSON tree transformation, we know that $|T_C| \geq |T_S|$. Then, the mapping size is the size of the smaller tree $|T_S|$ subtracted by the number of unmapped nodes in $T_S$.

Further, there are at most $4(n-k) + 2(3k) + 1$ nodes with identical labels. 1 results from the identical object in the root nodes, $s_1, \ldots, s_m$ and their child literals $L$ may be identical to $c_{i_j}$ and their child literals, and all other subtrees ($n-k$ many) share one object node and three literals $L$.

The cost of the minimal JSON edit mapping $\gamma(M_{DNJ})$ between $T_S$ and $T_C$ is bounded as follows:

$$\gamma(M_{DNJ}) \geq (|M_{DNJ}| - (4(n-k) + 2(3k) + 1))$$
$$+ (|T_C| - |M_{DNJ}|) + (|T_S| - |M_{DNJ}|)$$
$$\gamma(M_{DNJ}) \geq 4n - 4k - d + (|T_C| - |M_{DNJ}|) + (|T_S| - |M_{DNJ}|)$$
$$\gamma(M_{DNJ}) \geq 4n - 2k + (|T_S| - |M_{DNJ}|)$$
$$\gamma(M_{DNJ}) \geq 4n - 2k + d$$
$$\gamma(M_{DNJ}) \geq 4n - 2k$$

Note that the first term is an upper bound on the number of rename operations, the second term is the number of deletions, and the third term is the number of insertions.

Since the minimal JSON edit mapping is a visual representation of the node edit operations, we know that $\gamma(M_{DNJ}) = DNJEDI(T_S, T_C)$. Hence, $DNJEDI(T_S, T_C) \geq 4n - 2k$. □

LEMMA 6. *Given an instance of* X3C *with JSON tree transformations $T_S$ and $T_C$, the following holds: if there is an exact cover by 3-sets, then $DNJEDI(T_S, T_C) \leq 4n - 2k$.*

PROOF. Create a JSON edit mapping $M^*_{DNJ}$ as follows:

- Map nodes $s_1, s_2, \ldots, s_m$ in $T_S$ to a $c_{i_j}$ in $T_C$ with the same label. There must be zero-cost mappings for all $s_i \in S$ as well as their literal children $L$, since there exists an exact cover by 3-sets. Hence, for each triplet of $\lambda$-subtrees ($k$ many), 2 nodes have to be inserted ($l_j$ and an object node). The resulting costs are $2k$.
- Map the remaining $\mu$-subtrees ($n-k$ many) in $T_S$ to the remaining $\kappa$-subtrees in $T_C$ with cost 4.

The cost of the resulting mapping is $\gamma(M^*_{DNJ}) = 2k + 4(n-k) = 4n - 2k$. Since $M^*_{DNJ}$ is a valid but not necessarily minimal JSON edit mapping, $\gamma(M^*_{DNJ}) \geq DNJEDI(T_S, T_C)$. □

LEMMA 7. *Given an instance of* X3C *with JSON tree transformations $T_S$ and $T_C$, the following holds: if $DNJEDI(T_S, T_C) \leq 4n - 2k$, then there is an exact cover by 3-sets.*

PROOF. From $DNJEDI(T_S, T_C) \leq 4n - 2k$ and Lemma 5, we know that $DNJEDI(T_S, T_C) = 4n - 2k$. Further, from Lemma 5, we know that $d = 0$, hence all nodes in $T_S$ are in the minimal JSON edit mapping $M_{DNJ}$. We continue by analyzing which nodes pairs of $T_S$ and $T_C$ are in $M_{DNJ}$.

First, the root of $T_S$ has to be mapped to the root of $T_C$. Otherwise, the root can only be mapped to an object in a $\kappa$-subtree. Due to the ancestor condition, all nodes in $T_S$ have to be mapped to the subtree of the according object node in $T_C$. Since there are more nodes in $T_S$ than in such a subtree, this contradicts the fact that all nodes in $T_S$ are mapped.

Next, consider the root node of any $\mu$-subtree. Due to the type constraint of the JSON edit mapping, it can only be mapped to the root of a $\kappa$-subtree or to a $c_{ij}$. Mapping it to the root implies a minimal cost of 4 rename operations compared to at least 6 deletion operators otherwise. Therefore, in a minimal mapping, each $\mu$-subtree is mapped to a $\kappa$-subtree with cost 4, which results in an overall cost of $4(n-k)$.

Hence, only $2k$ edit operations remain to map $3k$ $\lambda$-subtrees to $k$ $\kappa$-subtrees. The size difference between these remaining subtrees in $T_C$ and $T_S$ is $8k - 6k = 2k$. Hence, all nodes in $T_S$ must be mapped with cost 0, i.e., all $s_i \in S$ must match exactly one $c_{ij}$ in $k$ $\kappa$-subtrees

JEDI: These aren't the JSON documents you're looking for... (Extended Version*)

in $T_C$, which is equivalent to $C'$ and , therefore, there is exists an exact cover by 3-sets. □

Theorem 5. *Computing the JSON edit distance between two JSON trees is NP-complete.*

Proof. We need to show the following two steps:
- $DNJEDIx \in NP$: A non-deterministic algorithm can guess a JSON edit mapping $M^*_{DNJ} \subseteq N(T_S) \times N(T_C)$ between two JSON trees $T_S$ and $T_C$ and verify in polynomial time whether the mapping costs are at most $i$.
- X3C can be reduced to DNJEDIx: From Lemma 6 and Lemma 7, it immediately follows that there is an exact cover by 3-sets iff $DNJEDI(T_C, T_S) \leq 4n - 2k$.

Hence, the computation of the JSON edit distance is NP-complete. □

While the above proof only holds for JSON trees with degree 3, the proof for trees with degree $deg = q > 3$ results from an according reduction of exact cover by $q$-sets. In this general case, the claim is as follows: there exists an exact cover by $q$-sets iff $DNJEDI(T_S, T_C) \leq (q+1)n + (q-5)k$.

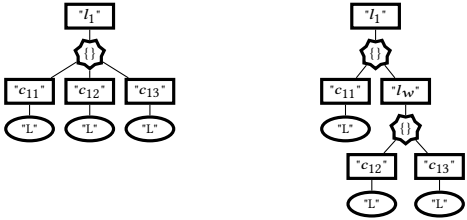

Figure 14: Transformation of $\kappa$- and $\mu$-subtrees into a JSON tree with degree 2, $l_w \notin S$.

Since exact cover by 2-sets is solvable in polynomial time, we reduce exact cover by 3-sets to show the NP-completeness of trees with degree two. The following two modifications are needed: (1) compared to Figure 13, adjust the structures of $\kappa$- and $\mu$-subtrees as shown in Figure 14, (2) the sizes change as follows $|T_C| = 10n + 1$, $|T_S| = 10(n - k) + 2(3k) + 1$, $|M| = 10n - 4k + 1 - d$, and there are at most $5(n - k) + 2(3k) + 1$ nodes with identical labels. Then it holds that $DNJEDI(T_C, T_S) \leq 5n - 3k$ iff there is an exact 3-cover. Hence, the JSON edit distance is NP-complete for JSON trees with degree 2 or higher. Note that DNJEDI for trees with degree 1 can be computed using the edit distance. This observations follows from the ancestor condition of DNJEDI.

Summarizing, the high computational complexity of DNJEDI disqualifies its application in real-world scenarios where JSON trees with thousands of nodes need to be compared.

## B REAL-WORLD JSON DATASET ANALYSIS

In Table 3, we give a summary of the characteristics of the datasets used in Section 6.

---

[5]https://www.kaggle.com/Cornell-University/arxiv
[6]https://mtgjson.com/api/v5/AtomicCards.json
[7]https://www.kaggle.com/dataturks/clothing-item-detection-for-ecommerce
[8]https://www.kaggle.com/fda/fda-enforcement-actions?select=food-enforcement-0001-of-0001.json
[9]https://www.kaggle.com/dataturks/face-detection-in-images
[10]https://data.nasa.gov/resource/y77d-th95.json
[11]https://www.kaggle.com/gjbroughton/christmas-recipes
[12]https://www.reddit.com/r/science.json
[13]https://www.kaggle.com/rtatman/the-national-university-of-singapore-sms-corpus
[14]https://social-dynamics.net/pepmusic/musicdata.zip
[15]https://www.kaggle.com/stanfordu/stanford-question-answering-dataset
[16]https://developer.twitter.com/en/docs.html

| | | Arxiv[5] | | | Cards[6] | | | Clothing[7] | | | DBLP [15] | | | DENF[8] | | | Device[8] | | |
|---|---|---|---|---|---|---|---|---|---|---|---|---|---|---|---|---|---|---|---|
| | #records | | 1,865,501 | | | 20,746 | | | 504 | | | 1,984,049 | | | 7,497 | | | 147,899 | |
| | #nodes/rec | 38 | 53 | 11367 | 37 | 132 | 298 | 29 | 29 | 51 | 14 | 26 | 651 | 47 | 59 | 283 | 49 | 323 | 3264 |
| | #objects | 2 | 2.6 | 188 | 3 | 12 | 39 | 4 | 4 | 7 | 2 | 2 | 2 | 2 | 2 | 2 | 2 | 2 | 2 |
| | #arrays | 3 | 6.2 | 2834 | 8 | 8 | 11 | 3 | 3 | 5 | 0 | 1.5 | 5 | 0 | 2.7 | 20 | 0 | 1.9 | 2 |
| | #keys | 16 | 17.1 | 388 | 17 | 59 | 127 | 12 | 12 | 21 | 6 | 11 | 17 | 23 | 27 | 45 | 24 | 29.9 | 39 |
| | #literals | 17 | 28 | 8512 | 9 | 54 | 192 | 10 | 10 | 18 | 5 | 11.5 | 634 | 22 | 28 | 221 | 23 | 289 | 3230 |
| object degree | min | 2 | 2 | 2 | 0 | 0.9 | 1 | 2 | 2 | 2 | 1 | 1 | 1 | 0 | 2.7 | 20 | 0 | 5.9 | 6 |
| | avg | 2 | 7.1 | 8 | 3 | 5.2 | 12 | 3 | 3 | 3 | 3 | 5.6 | 8.5 | 12 | 13.4 | 23 | 12 | 14.9 | 15 |
| | max | 14 | 14 | 14 | 16 | 21.5 | 28 | 5 | 5 | 5 | 5 | 10 | 16 | 23 | 24 | 25 | 24 | 24 | 24 |
| array degree | min | 1 | 1.4 | 4 | 0 | 0.03 | 1 | 1 | 1 | 1 | 0 | 1.5 | 263 | 0 | 0.2 | 1 | 0 | 131 | 1600 |
| | avg | 1.7 | 2.6 | 63.7 | 0.3 | 1.7 | 21 | 1.3 | 1.3 | 1.6 | 1 | 1.9 | 312 | 0 | 0.3 | 11 | 0 | 131 | 1602 |
| | max | 3 | 5 | 2832 | 1 | 7 | 167 | 2 | 2 | 2 | 1 | 2.3 | 621 | 0 | 1.3 | 95 | 0 | 131 | 1603 |
| | max. depth | 5 | 5 | 5 | 5 | 6.9 | 7 | 8 | 8 | 8 | 4 | 4 | 4 | 3 | 3.3 | 5 | 3 | 5 | 5 |
| | | Face[9] | | | FENF[8] | | | Movie[40] | | | NASA[10] | | | NBA [15] | | | Reads[10] | | |
| | #records | | 409 | | | 13,991 | | | 8,765,568 | | | 1,000 | | | 31,686 | | | 30,000 | |
| | #nodes/rec | 29 | 68 | 359 | 47 | 49 | 51 | 15 | 23 | 729 | 15 | 27 | 31 | 659 | 977 | 1269 | 11 | 11 | 11 |
| | #objects | 4 | 9.3 | 49 | 2 | 2 | 2 | 1 | 1.9 | 2 | 1 | 2 | 2 | 21 | 27 | 35 | 1 | 1 | 1 |
| | #arrays | 3 | 6.5 | 33 | 0 | 0 | 0 | 0 | 0 | 2 | 0 | 1 | 1 | 3 | 3 | 3 | 0 | 0 | 0 |
| | #keys | 12 | 28 | 147 | 23 | 24 | 25 | 7 | 11 | 13 | 7 | 12 | 14 | 339 | 477 | 619 | 5 | 5 | 5 |
| | #literals | 10 | 24 | 130 | 22 | 23 | 24 | 7 | 10 | 714 | 7 | 12 | 14 | 332 | 470 | 612 | 5 | 5 | 5 |
| object degree | min | 2 | 2 | 2 | 0 | 0 | 0 | 1 | 1.4 | 11 | 2 | 2.1 | 7 | 1 | 1 | 1 | 5 | 5 | 5 |
| | avg | 3 | 3 | 3 | 11.5 | 12 | 12.5 | 4.5 | 5.8 | 11 | 5.5 | 6.1 | 7 | 16 | 17.4 | 18.4 | 5 | 5 | 5 |
| | max | 5 | 5 | 5 | 23 | 24 | 25 | 7 | 10 | 12 | 7 | 10.2 | 12 | 20 | 20.5 | 21 | 5 | 5 | 5 |
| array degree | min | 1 | 1 | 1 | 0 | 0 | 0 | 0 | 0 | 704 | 0 | 1.9 | 2 | 2 | 2 | 2 | 0 | 0 | 0 |
| | avg | 1.3 | 1.6 | 1.9 | 0 | 0 | 0 | 0 | 0 | 704 | 0 | 1.9 | 2 | 5.3 | 7.5 | 10 | 0 | 0 | 0 |
| | max | 2 | 3.1 | 16 | 0 | 0 | 0 | 0 | 0 | 704 | 0 | 1.9 | 2 | 7 | 10.7 | 14 | 0 | 0 | 0 |
| | max. depth | 8 | 8 | 8 | 3 | 3 | 3 | 2 | 3.9 | 4 | 2 | 5 | 5 | 8 | 8 | 8 | 2 | 2 | 2 |
| | | Recipes[11] | | | Reddit[12] | | | Schema[5] | | | SMSen[13] | | | SMSzh[13] | | | Spotify[14] | | |
| | #records | | 1,617 | | | 25 | | | 81,518 | | | 55,835 | | | 31,465 | | | 1,107 | |
| | #nodes/rec | 13 | 26.1 | 64 | 209 | 265 | 432 | 3 | 270.5 | 48,128 | 81 | 81 | 81 | 79 | 80.9 | 81 | 5,141 | 5,141 | 5,146 |
| | #objects | 1 | 1 | 1 | 5 | 13 | 27 | 1 | 51.7 | 7,680 | 19 | 19 | 19 | 19 | 19 | 19 | 2 | 2 | 2 |
| | #arrays | 2 | 2 | 2 | 6 | 7.8 | 10 | 0 | 11.4 | 1,454 | 0 | 0 | 0 | 0 | 0 | 0 | 5 | 5 | 5 |
| | #keys | 6 | 6 | 6 | 104 | 130 | 206 | 1 | 122 | 23,768 | 40 | 40 | 40 | 39 | 39.9 | 4,006 | 19 | 19 | 19 |
| | #literals | 7 | 17.1 | 55 | 94 | 115 | 189 | 1 | 85.4 | 16,307 | 22 | 22 | 22 | 21 | 21.9 | 22 | 5,115 | 5,115 | 5,120 |
| object degree | min | 6 | 6 | 6 | 0 | 0 | 0 | 0 | 1.2 | 10 | 1 | 1 | 1 | 1 | 1.4 | 11 | 9 | 9 | 9 |
| | avg | 6 | 6 | 6 | 7.6 | 11.3 | 20.8 | 1 | 2.7 | 10 | 2.1 | 2.1 | 2.1 | 4.5 | 5.8 | 11 | 9.5 | 9.5 | 9.5 |
| | max | 6 | 6 | 6 | 102 | 103 | 104 | 1 | 13.2 | 531 | 9 | 9 | 9 | 7 | 10 | 12 | 10 | 10 | 10 |
| array degree | min | 0 | 3.5 | 19 | 0 | 0 | 0 | 0 | 1.4 | 463 | 0 | 0 | 0 | 0 | 0 | 0 | 1 | 1 | |
| | avg | 1.5 | 6.5 | 25.5 | 0 | 0.7 | 1.9 | 0 | 2.2 | 745.5 | 0 | 0 | 0 | 0 | 0 | 0 | 1,020 | 1,020 | 1,021 |
| | max | 2 | 9.6 | 41 | 0 | 3.9 | 6 | 0 | 4.7 | 2,810 | 0 | 0 | 0 | 0 | 0 | 0 | 3,600 | 3,600 | 3,600 |
| | max. depth | 3 | 3 | 3 | 5 | 10.9 | 11 | 2 | 11.6 | 50 | 8 | 8 | 8 | 8 | 8 | 8 | 6 | 6 | 6 |
| | | StanDev[15] | | | StanTrain[15] | | | Twitter 2[16] | | | Virus[10] | | |
| | #records | | 48 | | | 442 | | | 19,316 | | | 500 | | |
| | #nodes/rec | 2212 | 5379 | 18,135 | 294 | 2597 | 10,549 | 115 | 195 | 489 | 19 | 19 | 19 |
| | #objects | 404 | 988 | 3303 | 50 | 440 | 1783 | 5 | 11.4 | 42 | 1 | 1 | 1 |
| | #arrays | 118 | 264 | 864 | 28 | 242 | 966 | 2 | 13.2 | 27 | 0 | 0 | 0 |
| | #keys | 904 | 2196 | 7416 | 122 | 1078 | 4383 | 55 | 83 | 222 | 9 | 9 | 9 |
| | #literals | 786 | 1931 | 6551 | 94 | 837 | 3417 | 53 | 87 | 208 | 9 | 9 | 9 |
| object degree | min | 2 | 2 | 2 | 2 | 2 | 2 | 0 | 1.5 | 2 | 9 | 9 | 9 |
| | avg | 2.1 | 2.2 | 2.3 | 2.4 | 2.45 | 2.5 | 4.4 | 7.75 | 11.9 | 9 | 9 | 9 |
| | max | 3 | 3 | 3 | 3 | 3 | 3 | 29 | 34 | 35 | 9 | 9 | 9 |
| array degree | min | 1 | 2.2 | 4 | 1 | 1 | 1 | 1 | 1 | 2 | 0 | 0 | 0 |
| | avg | 3.1 | 3.7 | 5.3 | 1.7 | 1.8 | .9 | 1.7 | 2 | 2.5 | 0 | 0 | 0 |
| | max | 21 | 43 | 98 | 5 | 43 | 149 | 2 | 4.4 | 16 | 0 | 0 | 0 |
| | max. depth | 11 | 11 | 11 | 11 | 11 | 11 | 5 | 9.2 | 14 | 3 | 3 | 3 |

Table 3: Analysis of the collected JSON datasets.